\documentclass[11pt]{article}

\usepackage[final]{acl}
\usepackage{times}
\usepackage{latexsym}
\usepackage[T1]{fontenc}
\usepackage[utf8]{inputenc}
\usepackage{microtype}
\usepackage{inconsolata}
\usepackage{graphicx}
\usepackage{enumitem}
\usepackage[most]{tcolorbox}
\usepackage{algorithm}
\usepackage{algpseudocode}
\usepackage{mdframed}
\usepackage{multirow}
\usepackage{xspace}
\usepackage{hyperref}
\usepackage{tikz}
\usepackage{listings}
\usepackage{xcolor}
\usepackage{caption}
\usepackage{array}
\usepackage{booktabs}
\usepackage[flushmargin]{footmisc}
\usepackage{cleveref}

\makeatletter
\renewcommand{\theHALG@line}{\thealgorithm.\arabic{ALG@line}}
\makeatother

\newcommand{\matex}{$\textit{MATEX}$\xspace}
\newcommand{\txsum}{$\textit{TxSum}$\xspace}

\newcommand{\mypara}[1]{\smallskip\noindent{\bf {#1}.}\xspace}

\title{\txsum: User-Centered Ethereum Transaction Understanding with Micro-Level Semantic Grounding}

\author{
\textbf{Zifan Peng}\textsuperscript{1,2} \quad
\textbf{Jingyi Zheng}\textsuperscript{1} \quad
\textbf{Yule Liu}\textsuperscript{1} \quad
\textbf{Huaiyu Jia}\textsuperscript{1} \quad
\textbf{Qiming Ye}\textsuperscript{1} \quad
\textbf{Jingyu Liu}\textsuperscript{1} \\
\textbf{Xufeng Yang}\textsuperscript{\textbf{3}} \quad
\textbf{Mingchen Li}\textsuperscript{\textbf{4}} \quad
\textbf{Qingyuan Gong}\textsuperscript{\textbf{5}} \quad
\textbf{Xuechao Wang}\textsuperscript{\textbf{1}} \quad
\textbf{Xinlei He}\textsuperscript{\textbf{2}}\thanks{Corresponding author: Xinlei He (xinlei.he@whu.edu.cn).} \\
\textsuperscript{1}The Hong Kong University of Science and Technology (Guangzhou) \quad
\textsuperscript{2}Wuhan University \\
\textsuperscript{3}China National Tobacco Corporation Liaoning Province Company \\
\textsuperscript{4}University of North Texas \quad
\textsuperscript{5}Fudan University
}

\begin{document}

\maketitle

\begin{abstract}
Understanding the economic intent of Ethereum transactions is critical for user safety, yet current tools expose only raw on-chain data or surface-level intent, leading to widespread ``blind signing'' (approving transactions without understanding them).
Through interviews with 16 Web3 users, we find that effective explanations should be structured, risk-aware, and grounded at the token-flow level.
Motivated by these findings, we formulate \txsum, a new domain-grounded NLP task for DeFi transaction explanation, and construct a dataset of 187 complex Ethereum transactions with 2,375 token-flow annotations and transaction-level summaries.

We further introduce \matex, a grounded multi-agent framework for high-stakes transaction explanation.
It selectively retrieves external knowledge under uncertainty and audits explanations against raw traces to improve token-flow-level factual consistency.
\matex achieves the strongest overall explanation quality, especially on micro-level factuality and intent quality.
It improves user comprehension on complex transactions from 52.9\% to 76.5\% over the strongest baseline and raises malicious-transaction rejection from 36.0\% to 88.0\%, while maintaining a low false-rejection rate on benign transactions.
\end{abstract}

\section{Introduction}
\label{sec:intro}

Understanding the economic intent of an Ethereum transaction is crucial for both proactive risk mitigation and post-incident analysis.
It enables users to preview outcomes and avoid irreversible errors before signing, and supports regulatory compliance, anti-money laundering, and forensic diagnosis after exploits—such as the TraceLLM~\citep{wang2025tracellmsecuritydiagnosistraces} case, where on-chain analysis uncovered attack vectors in major DeFi incidents.

However, this understanding is critically lacking in practice.
Modern DeFi protocols can coordinate multiple contracts and even multiple chains, as illustrated by bridgeless cross-chain options~\citep{peng2025crosschain}; their resulting transactions are highly compositional, making correct human review difficult and contributing to ``blind signing''~\citep{liu2024defi,blindsigning}—a phenomenon with severe consequences.
In early 2025, attackers stole \$1.5B from Bybit by tricking operators into approving a malicious transaction that appeared innocuous in their interface~\cite{bybit2025attack}.
In routine interactions, new users struggle with dual-sign patterns (approve + swap), while experienced users rely on “reject unlimited approvals”, a heuristic that fails across contracts.

These incidents reveal a fundamental gap: current tools expose raw on-chain data but fail to \textbf{translate it into human-understandable economic narratives}.
While tools like MetaSuites~\cite{metasuites}, EigenPhi~\citep{eigenphi}, and Tenderly~\citep{tenderly} expose detailed token flows and contract interactions, they stop at the syntactic level.
They expose detailed execution artifacts rather than synthesizing what these low-level events mean economically for users.

\begin{figure*}[hbtp]
    \centering
    \includegraphics[width=0.9\linewidth]{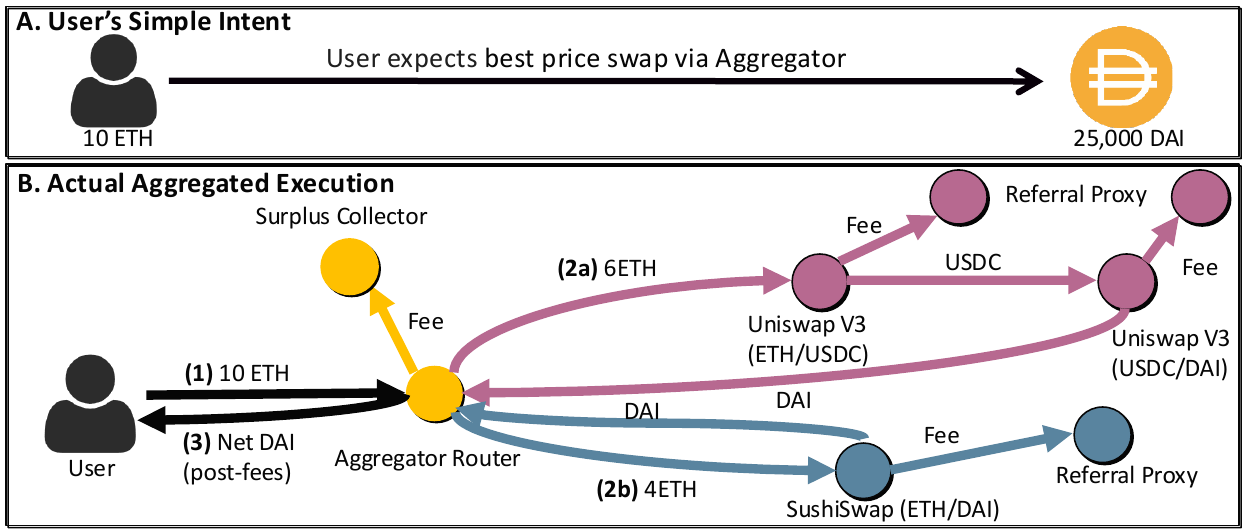}
    \caption{Mismatch between user intent and on-chain execution.
    A simple swap intent is realized through a complex and opaque transaction trace.
    The example is simplified for clarity.}
    \label{fig:flow-ex}
\end{figure*}

As \Cref{fig:flow-ex} shows, a user’s simple intent (e.g., “swap ETH for DAI”) often spans multiple contracts.
Raw traces show \textit{what} happened but not \textit{why}, leaving users unable to distinguish a routine swap from a high-risk operation.
Three research questions guide our study:

\begin{itemize}[leftmargin=*, itemsep=0pt, topsep=0pt]
    \item \textbf{RQ1:} How do Web3 users understand transaction details when signing?
    \item \textbf{RQ2:} What do users expect from transaction explanations in terms of content and format?
    \item \textbf{RQ3:} How to design a framework and how the generated explanations by it affect users’ comprehension and signing behavior?
\end{itemize}

To address this gap, we conducted semi-structured interviews with 16 Web3 users spanning traders, developers, and casual users.
Three findings directly shaped our formulation: (1) blind signing is widespread even among experienced users; (2) existing tools expose syntactic traces but not user-facing economic semantics; and (3) users naturally reason at the token-flow level while still expecting a concise transaction-level summary.
Full study details are provided in Appendix~\ref{app:interview_protocol}.

\begin{figure*}[t]
    \centering
    \includegraphics[width=1.0\linewidth]{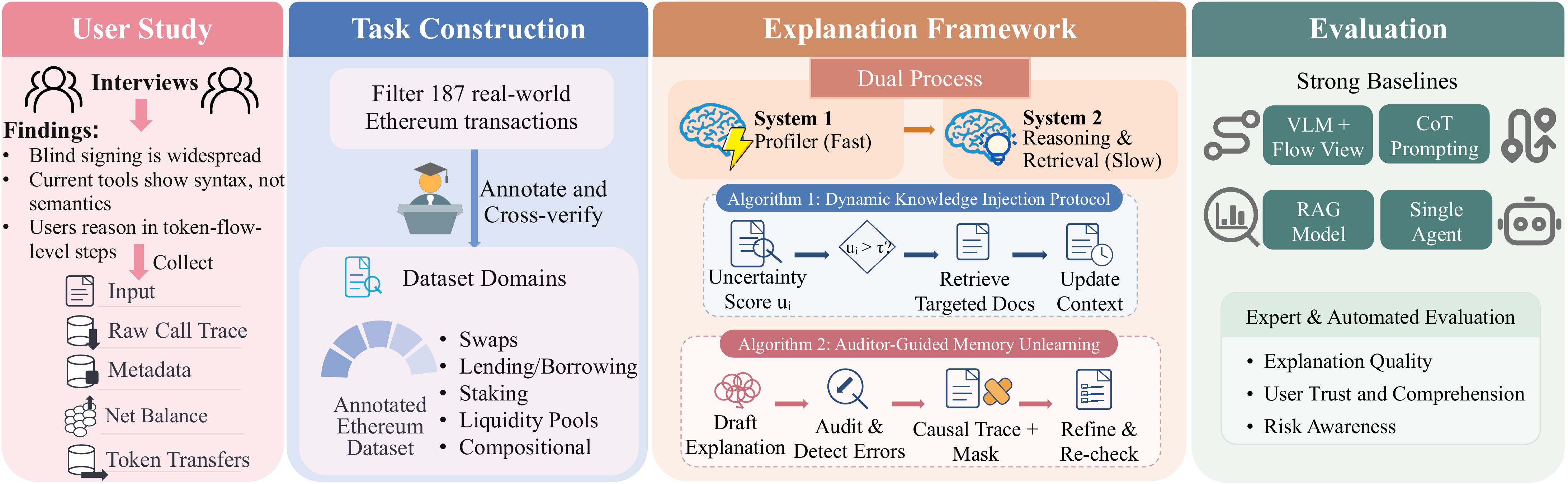}
    \caption{Overview of our study.
    We first conduct user interviews to identify the explanatory needs behind blind signing, which motivate \txsum, a new domain-grounded NLP task and dataset for DeFi transaction explanation.
    We then introduce \matex, a grounded multi-agent method for this task, and evaluate it against strong baselines using automatic, expert, and user-centered assessments.}
    \label{fig:teaser}
\end{figure*}

Guided by these findings, we formulate the paper around one central thesis: we introduce a new domain-grounded NLP task for DeFi transaction explanation and a grounded method for improving micro-level factual explanation under high-stakes reasoning.
As illustrated in Figure~\ref{fig:teaser}, our study proceeds from user interviews, to task and dataset construction, to grounded explanation modeling, and finally to multi-faceted evaluation.

Concretely, we propose \txsum, a task and schema for transaction understanding at the token-flow level.
\txsum defines 5 semantic attributes (see~\Cref{sec:understand}) that ground explanations in step-wise, verifiable actions aligned with user mental models.
We then construct a dataset of 187 Ethereum transactions containing 2,375 token-flow annotations.
Each transaction includes (1) a natural-language summary and (2) fine-grained flow-level labels following our predefined schema.

We further introduce \matex, a grounded multi-agent framework motivated by how human experts analyze transactions.
Rather than using unrestricted tool-augmented generation, \matex implements a dual-process reasoning workflow~\citep{booch2021thinking}: fast pattern recognition (System~1) flags uncertainty and triggers targeted evidence-seeking investigation (System~2), followed by adversarial factual auditing against raw traces.
The design explicitly targets the core NLP challenge in this setting: generating faithful, micro-level explanations for complex DeFi traces without hallucinating unseen protocol behavior.

We evaluate \matex through automatic, expert, and user studies.
Results show that \matex consistently outperforms strong baselines---including monolithic prompting, VLM prompting with flow visualization, a single-agent tool user, Few-shot CoT, and a RAG pipeline---in factuality, intent quality, and risk coverage (see \Cref{sec:evaluation}).
In summary, we make the following contributions:

\begin{itemize}[leftmargin=*, itemsep=0pt, topsep=0pt]
    \item We identify user-centered explanatory needs through semi-structured interviews with 16 diverse Web3 participants and define a structured schema for transaction understanding grounded in real usage.

    \item We introduce \txsum, a new domain-grounded NLP task and dataset for DeFi transaction explanation, pairing natural-language summaries with 2,375 step-wise semantic annotations across 187 complex, real-world transactions.

    \item We propose \matex\footnote{\url{https://github.com/sfofgalaxy/matex}}, a grounded multi-agent framework that combines uncertainty-triggered retrieval, evidence synthesis, and adversarial factual auditing to generate accurate, risk-aware explanations from raw on-chain data, significantly improving user comprehension and reducing ``blind signing'' across automatic, expert, and user evaluations.
\end{itemize}

\section{Understanding Transaction Flow}
\label{sec:understand}
Our user interviews directly motivated \txsum's schema.
While concurrent work~\citep{mao2025know} focuses on transaction-level intent mining over a predefined taxonomy, \txsum targets user-centered transaction understanding at two levels: (1) a concise transaction summary that explains the overall intent and outcome, and (2) token-flow annotations that capture step-wise economic actions.
This grounds micro-level explanations in user-relevant actions and risks.

\mypara{Task Definition}
\txsum is the task of generating a faithful explanation of a DeFi transaction from strictly on-chain observable data.
The input consists of four components:
(1) transaction metadata (e.g., sender, value, timestamp),
(2) the full call trace,
(3) a sequence of token transfers (ERC-20/721), and
(4) net balance changes for all involved accounts.

The output comprises two parts: (1) A transaction-level summary: a 3-4 sentence natural-language narrative that integrates the economic purpose, execution pathway, and outcomes of the transaction, including critical non-token state changes such as approvals, ownership transfers, or admin control updates.
(2) Token-flow annotations: for each token transfer, five semantic attributes—action\_type, intent, mechanism, precondition, and result—that ground the summary in step-wise, verifiable actions.
The full specification of the five attributes is detailed in~\Cref{sec:identified-attributes}.
The two output levels are complementary: flow annotations make each economic action checkable, while the summary preserves the concise view requested in our interviews.
Requiring both also prevents a fluent transaction-level narrative from masking unsupported intermediate claims.

The task formulation is broadly applicable to DeFi transaction explanation, but the current schema and implementation assume EVM-style traces and token flows.
EVM-compatible chains such as Arbitrum, Optimism, and Base require only limited tooling adaptation, whereas non-EVM chains such as Solana require changes to both the input adapters and parts of the semantic schema.

\subsection{Identified Attributes}
\label{sec:identified-attributes}
User interviews reveal that effective explanations must articulate the economic logic behind each token flow.
To formalize this, we define five semantic attributes for every flow:

\begin{enumerate}[leftmargin=*, itemsep=0pt, topsep=0pt]
    \item action\_type, a standardized label for one of 14 financial primitives (e.g., \textit{redeem}, \textit{deposit}) (defined in~\Cref{tab:action-taxonomy} in Appendix);
    \item intent, the user’s high-level purpose (e.g., ``swap out USDT for USDC'');
    \item mechanism, the technical pathway indicating which contract and function were called (e.g., ``UniswapV3\#exactInput'');
    \item precondition, the on-chain conditions required for success (e.g., USDC approved and balance > 10); and
    \item result, the factual state change observed after execution (e.g., +1500 USDC balance).
\end{enumerate}

These attributes turn raw transaction traces into explanations that are both easy to follow and explicit about risks, and that are aligned with user mental models.
Detailed definitions and examples are provided in Appendix~\ref{app:attributes}.

\subsection{TxSum Dataset}

\mypara{Data Collection}
To build a dataset that reflects real-world DeFi complexity while remaining feasible for expert annotation, we first sampled 500 Ethereum transactions from a full node over 2025.
Specifically, we sampled 100 transactions from each of five expert-defined semantic categories: swaps (including multi-hop and aggregator-routed), lending/borrowing cycles, staking and delegation, liquidity pool operations, and compositional or cross-protocol interactions (e.g., flash loans, arbitrage, or custom logic not fitting the other categories).
Because DeFi transactions often combine behaviors, we use these categories only for sampling and non-exclusive coverage profiling, not to impose a single label on each transaction.

We then filtered the 500 sampled transactions by token-flow length.
We removed transactions with fewer than 5 token flows because they were usually simple transfers or otherwise too easy for step-wise semantic annotation.
We also removed transactions with more than 30 token flows because they were too costly to annotate reliably.
After filtering, the final \txsum dataset contains 187 transactions.
For each transaction, we collect transaction metadata and the full call trace from an Ethereum archival node, ordered token transfers and net balance changes from EigenPhi~\citep{eigenphi}, and the token-flow graph from EigenPhi.

\mypara{Data Annotation}
The 187 transactions were annotated by six graduate-level annotators whose research focuses on DeFi.
Annotators were organized into three fixed pairs, and each pair independently annotated a disjoint subset of 62 or 63 transactions.
For every transaction, the two annotators in the assigned pair independently completed the full annotation, including token-flow labels and the transaction-level summary.
Any disagreements were then resolved by a third annotator.
Our annotation pipeline mirrors the cognitive process users described in interviews: (1) understand token movement, (2) infer purpose and mechanism, and (3) synthesize a coherent narrative.
The process consists of three steps:

\noindent \textbf{Step 1: Transaction Overview.}
Annotators first review all collected data to understand the transaction’s purpose.
They then draft a coarse summary to guide labeling; it is excluded from the dataset.

\noindent \textbf{Step 2: Semantic Labeling.}
For each token transfer, annotators label five attributes derived from user interviews:
action\_type, intent, mechanism, precondition, and result.

\noindent \textbf{Step 3: Narrative Synthesis.}
Annotators write a 3-4 sentence natural-language summary that strictly reflects observable evidence, without invoking external knowledge, protocol reputations, or speculative intent.
Detailed annotation process and examples are provided in Appendix~\ref{app:annotation}.

\mypara{Data Statistics}
The final dataset contains 187 annotated transactions and 2,375 token-flow annotations, with an average of 12.7 token-flow steps per transaction.
It focuses on compositionally complex DeFi interactions rather than routine transfers or short single-step executions.
The resulting summaries average 3.3 sentences (101.3 tokens).
Based on a heuristic of protocol and control-flow complexity, we categorize 43 transactions as simple (single protocol), 90 as medium (2--3 protocols), and 54 as complex ($\geq 4$ protocols or nested logic).
The five sampled scenarios cover swaps, lending/borrowing, staking, liquidity-pool operations, and cross-protocol composition, which align with major DeFi sectors identified by both current ecosystem taxonomies and prior protocol surveys~\citep{defillama_categories,shah2023defi_review}.

Moreover, 20.2\% of all contracts involved in the dataset lack a verified ABI, so the evaluation includes interactions for which protocol semantics cannot be recovered from verified source code alone.
Full statistics are shown in Appendix~\ref{app:dataset_profile}.
To assess annotation consistency at the token-flow level, we evaluate the agreement of annotators.
For action\_type, we compute inter-annotator agreement using Cohen's $\kappa$, yielding action\_type = 0.98.

For the natural language fields intent, mechanism, precondition, and result, exact-form agreement would underestimate reliability because annotators may produce semantically equivalent paraphrases.
We therefore report pairwise semantic agreement using BERTScore F1, computed between the two independently written annotations for each token flow and averaged over the dataset.
Specifically, for each token flow, we compute BERTScore F1 between the two independently written annotations using the official BERTScore~\citep{bertscore} implementation with a roberta-large backbone and baseline rescaling enabled, and then average the resulting scores over the dataset; all four fields score above 0.91 (see details in~\Cref{tab:semantic_consistency} in Appendix).
These results indicate high consistency across attributes, supporting the reliability of the dataset.

\section{Multi-Agent Framework: \textit{MATEX}}
\label{sec:matex}

To address \textbf{RQ3} (how to explain it), we start from how human experts analyze complex transactions.
They do not rely on memorized patterns alone; instead, they follow a clear workflow:
(1) quickly scan token flows for familiar patterns,
(2) stop when something is unclear,
(3) look up documentation or code,
(4) synthesize the evidence into an explanation, and
(5) check for hidden risks.

\mypara{Dynamic Knowledge Injection Protocol}
The key decision is when to move from System~1 to System~2.
Most token flows are routine and can be interpreted from on-chain evidence alone.
Only a small number are ambiguous and need extra protocol context.
For this reason, \matex uses Dynamic Knowledge Injection Protocol (DKIP).
The Profiler computes a semantic uncertainty score $u_i$ for each token flow $t_i$.
If $u_i$ is low, the system keeps a lightweight explanation path.
If $u_i$ is above a threshold $\tau$, the Investigator retrieves protocol evidence such as ABI entries or verified documentation.
The Profiler also emits a retrieval key when escalation is needed.
This design avoids unnecessary retrieval and keeps the context focused on uncertain flows.

\matex does not require a verified contract or an available ABI.
When ABI evidence is missing, it falls back to observable token transfers, event logs, net balance changes, call-trace structure, and call direction; the resulting explanation uses conservative descriptions such as ``swap-like routing'' and emits an uncertainty warning instead of asserting unverifiable function semantics.

External information can be misleading or unreliable; prior prompt-based curriculum and contrastive methods therefore treat credibility assessment as an explicit prediction problem~\citep{peng2023promptcurriculum,peng2025promptcontrastive}.
Because AI-powered search can also surface malicious or unverified external content~\citep{luo2025unsafe}, retrieved context is not accepted as ground truth: the Safety Auditor checks resulting claims against raw traces.

\mypara{Auditor-Guided Context Pruning}
After retrieval, System~2 produces and checks the explanation.
The Synthesizer combines on-chain evidence, retrieved protocol context, and flow-level hypotheses to produce a draft summary.
The Safety Auditor then checks this draft against the raw trace.
Here \matex uses Auditor-Guided Context Pruning (AGCP).
The intuition is that unsupported claims can remain in the working context and affect later drafts.
Instead of only appending critique, AGCP identifies the error-causing context $P_{\mathrm{mask}}$ and removes it before the next refinement step.
This reduces repeated hallucinations across iterations.

This design is connected to the ``Lost in Conversation'' failure mode, in which language models over-rely on earlier incorrect attempts in a multi-turn context~\citep{laban2025lostconversation}.
AGCP acts as a structural mitigation for this form of context contamination because it removes unsupported intermediate content rather than merely appending a correction.
The full pseudocode of DKIP and AGCP are provided in Appendix~\ref{app:matex_algorithms}.

\begin{itemize}[leftmargin=*, itemsep=0pt, topsep=0pt]
    \item \textbf{Profiler} implements System~1.
    It scans token flows and computes a semantic uncertainty score $u_i$ for each token flow $t_i$.

    \item \textbf{Investigator} supports System~2 with retrieving live protocol information only when $u_i$ is above the threshold $\tau$, following DKIP.

    \item \textbf{Synthesizer} is the main reasoning module in System~2, which combines on-chain evidence, retrieved protocol context, and flow-level hypotheses to produce a transaction summary.

    \item \textbf{Safety Auditor} checks the draft explanation against the raw trace and applies AGCP when it finds unsupported facts or missing risks.
\end{itemize}

These components realize the full workflow of \matex:
System~1 decides whether a flow can be handled directly or should be escalated,
DKIP controls what evidence enters System~2,
the Synthesizer turns that evidence into a flow-grounded explanation,
and AGCP removes unsupported reasoning before the next draft.
This design supports our main goal: grounded micro-level factual explanation rather than only fluent transaction narration.
In our implementation, all agents use Qwen3-Max as the backbone LLM and are orchestrated with LangGraph~\citep{langgraph} to support stateful and iterative interactions.
The agents share structured tools for on-chain data access and protocol document retrieval.
Full implementation details are provided in Appendix~\ref{app:matex_details}.

\section{Evaluations and Results}
\label{sec:evaluation}

Our interviews reveal that existing interfaces fail to support safe DeFi interaction: users sign transactions they don’t understand, often missing latent risks.
To address \textbf{RQ3} (user comprehension and signing behavior), we give 3 evaluation objectives (EOs) that map directly to user-identified needs:

\begin{itemize}[leftmargin=*, itemsep=0pt, topsep=0pt]
    \item \textbf{EO1: Explanation Quality.}
    We evaluate the accuracy, intent coverage, outcome correctness, and clarity of generated summaries using automatic metrics and expert annotation.
    \item \textbf{EO2: User Comprehension.}
    We measure users' understanding, trust, and clarity via a controlled user study.
    \item \textbf{EO3: Risk Awareness.}
    We evaluate whether users can identify real-world security threats with the help of \matex.
\end{itemize}

\subsection{Baselines}
We compare \matex against five representative baselines:
\textbf{(1) Monolithic LLM.}
An LLM that processes the full input (trace, metadata) in one pass.
\textbf{(2) Few-shot CoT.}
A prompting baseline using the same Qwen3-Max backbone with 2 in-context demonstrations and an explicit chain-of-thought reasoning template.
\textbf{(3) VLM.}
A VLM augmented with a token-flow graph.
\textbf{(4) RAG.}
A retrieval baseline that extracts contract addresses, function signatures, and protocol names from the trace, retrieves external documents with a fixed workflow, and appends the retrieved evidence directly to the model.
\textbf{(5) Single-Agent.}
A ReAct-style agent using Qwen3-Max as its backbone, with access to the same web search and Etherscan tools as \matex.

To ensure a fair comparison, all text-only baselines (\matex, Monolithic LLM, Single-Agent, Few-shot CoT, Pure RAG) use the same Qwen3-Max~\citep{yang2025qwen3technicalreport} backbone, while the visualization baseline uses Qwen3-VL-Plus~\citep{bai2025qwen3vltechnicalreport} due to its multimodal input requirement.
All retrieval-enabled systems are granted the same retrieval permissions, tool budget, and temporal alignment: the same Etherscan API access, the same web-search scope, the same maximum number of external retrieval calls, and the same transaction-time protocol information window.
Single-Agent can call the same web search and Etherscan tools as \matex, while Pure RAG pipeline uses the same retrieval sources but without role specialization, iterative auditing, or uncertainty-triggered control.
This setup isolates the contribution of \matex's selective retrieval and auditing framework rather than confounding it with information asymmetry.

All systems receive identical on-chain inputs and output the result for the whole transaction.
We assess EO1 via automatic metrics and expert annotation, EO2 through a controlled user study, and EO3 using a user study on real attack transactions.
Full evaluation details are provided in Appendix~\ref{app:eval_details}.

\subsection{Evaluation Details and Results}
\label{sec:eval_metrics}

\subsubsection{EO1: Explanation Quality}
We assess transaction summaries using 4 binary (pass/fail) criteria:
\textbf{T1} (outcome correctness) and \textbf{T2} (clarity \& conciseness) at the transaction level, and
\textbf{F1} (faithfulness) and \textbf{F2} (intent quality) at the token-flow level; the full rubric is provided in~\Cref{tab:txsum-eval} in Appendix.

We use two evaluation methods:
(1) \textbf{LLM-as-judge}: GPT-5.1~\citep{gpt} scores all system outputs with binary decisions for each metric, and we explicitly validate the judge against expert labels;
(2) \textbf{Expert evaluation}: 3 DeFi Ph.D. researchers rate all transactions, assessing factual correctness, intent alignment, and risk awareness.
The resulting pass rates from both evaluation sources are reported in~\Cref{tab:eo1_results}.

For binary judgments, we additionally report ROUGE-L and BLEU for transaction-level summaries, Exact Match (EM) for flow-level semantic attribute generation, and Accuracy / Macro-F1 for the 14-way action\_type classification subtask.
As shown in Appendix~\ref{app:eo1_nlp}, \matex achieves the best performance on both summary-generation metrics (ROUGE-L, BLEU) and structured prediction metrics (EM, Accuracy, Macro-F1), consistent with the pass-rate results in~\Cref{tab:eo1_results}.

\begin{table*}[hbtp]
\centering
\caption{EO1 pass rates (\%): LLM-as-judge and expert human evaluation on baselines and \matex.}
\label{tab:eo1_results}
\resizebox{0.66\linewidth}{!}{
\begin{tabular}{l|cccc|cccc}
\toprule
\multirow{2}{*}{\textbf{Method}} & \multicolumn{4}{c|}{\textbf{Pass Rate (\%, LLM-as-judge)}} & \multicolumn{4}{c}{\textbf{Pass Rate (\%, Human)}} \\
& \textbf{T1} & \textbf{T2} & \textbf{F1} & \textbf{F2} & \textbf{T1} & \textbf{T2} & \textbf{F1} & \textbf{F2} \\
\midrule
LLM             & 79.1 & 80.7 & 51.2 & 64.3 & 71.2 & 82.1 & 53.4 & 65.7 \\
Few-Shot CoT    & 74.9 & 86.1 & 62.5 & 63.9 & 73.9 & 85.2 & 62.1 & 64.8 \\
VLM             & 77.0 & 82.9 & 60.8 & 61.2 & 75.8 & 86.0 & 61.4 & 62.8 \\
RAG             & 78.1 & 81.8 & 52.0 & 64.6 & 72.1 & 82.0 & 53.9 & 65.8 \\
Single-Agent    & 66.8 & 89.8 & 67.6 & 67.5 & 70.1 & 88.1 & 69.0 & 69.2 \\
\midrule
MATEX           & \textbf{73.3} & \textbf{93.0} & \textbf{73.2} & \textbf{76.6} & \textbf{76.0} & \textbf{94.0} & \textbf{75.1} & \textbf{78.3} \\
\bottomrule
\end{tabular}
}
\end{table*}

\mypara{Judge Reliability and Significance}
We validate the GPT-5.1 judge against expert annotations and find strong agreement overall, with per-metric Cohen's $\kappa$ and verbosity-bias checks reported in Appendix~\ref{app:judge_confusion}.
An additional Llama-3-70B judge yields the same ranking on F1/F2, indicating that the main conclusion is not specific to GPT-5.1.
Expert inter-rater reliability is also high (Fleiss' $\kappa=0.76$), supporting the reliability of the human evaluation protocol.

To assess robustness, we compare \matex against the strongest baseline using paired significance tests on the same 187 transactions: McNemar tests for binary pass/fail metrics and paired tests for ROUGE-L, EM, and action\_type Macro-F1.
The results confirm statistically significant gains on the key micro-level factuality and intent metrics.

\mypara{Results}
Results from both LLM-as-judge and expert human evaluation are summarized in~\Cref{tab:eo1_results}.
\matex achieves the highest scores on F1 and F2 across both evaluators, demonstrating its strength in grounded, step-wise reasoning.
On T1, \matex performs on par with the best baselines---scoring 76.0\% in human evaluation, the highest.
Manual inspection shows that \matex sometimes omits minor fees or implicit approvals absent from the call trace, consistent with its rule to report only facts verified on-chain.

The standard NLP metrics reinforce the same pattern: \matex achieves the best ROUGE-L, BLEU, Exact Match, Accuracy, and Macro-F1 under matched retrieval conditions.
This indicates that the gains are not confined to judge-based or human pass/fail assessment, but also transfer to standard generation and classification metrics.
Cross-backbone experiments with Llama-3-70B and Llama-3-8B show the same pattern: absolute scores fall with smaller open-weight models, but \matex consistently improves over the corresponding Single-Agent baseline on F1 and F2.

The performance patterns align with the architectural limitations of each baseline.
Both the Monolithic LLM and the VLM (augmented with a token-flow graph) process the input in a single pass without explicit decomposition or verification, making them prone to hallucination or missed details---especially on minor but critical flows such as small fees.
Few-shot CoT improves local reasoning structure but still lacks grounded verification.

The Pure RAG benefits from external evidence, yet its fixed retrieval-and-append workflow cannot selectively investigate ambiguous flows or audit generated claims against the raw trace.
The Single-Agent integrates decomposition, retrieval, synthesis, and narration into one cognitive loop, lacking role specialization or adversarial validation, which leads to error propagation in complex transactions.

\mypara{Efficiency and Cost}
\matex incurs higher latency than the Monolithic LLM and RAG baselines because it performs iterative reasoning and auditing, but it remains comparable to Single-Agent.
On complex transactions, \matex reduces average latency from 51.2 to 38.9 seconds and retrieval calls from 5.8 to 3.2 relative to Single-Agent; full latency, token, call, and estimated-cost profiles are reported in~\Cref{tab:cost_profile} in Appendix.

\subsubsection{EO2: User Trust \& Comprehension}
We conduct a separate controlled user study with a new cohort of 20 active DeFi users, distinct from the 16 interview participants (full details are provided in Appendix~\ref{app:eo2}).
Participants were recruited via Web3 communities and professional networks, and all reported at least three on-chain DeFi transactions in the past six months.
The cohort spans six countries/regions and includes 4 practitioners, 5 researchers/developers, 7 students, and 4 casual users.
All interfaces, explanations, and questionnaires were presented in English, so participants were required to be able to read English DeFi interfaces and transaction explanations.

EO2 uses a within-subject repeated-measures design where each participant views 9 transactions involving 113 token transfers.
For each level (simple, medium, and complex), participants see one transaction under each of the following conditions using the EigenPhi interface:
(a) Raw: the default explorer view only;
(b) Single-Agent: explorer view augmented with annotations from the strongest baseline in EO1;
(c) \matex: explorer view augmented with \matex explanations.
Transaction-condition assignment is balanced across participants using a Latin Square design.

\mypara{Results}
\Cref{tab:eo2_results} summarizes the EO2 results.
\matex consistently improves both objective comprehension and subjective trust, with the largest gains on complex transactions.
In particular, comprehension accuracy on complex transactions rises from 35.3\% in the Raw condition and 52.9\% in the Single-Agent condition to 76.5\% with \matex.
Perceived trust and clarity show a similar trend, increasing from 1.8--2.6 in the comparison conditions to 3.7--4.5 with \matex.

The improvements suggest \matex not only improves transaction understanding, but also supports safer signing decisions.
The advantage of \matex is especially pronounced on medium and high complexity transactions, where users benefit most from grounded, step-wise explanations.
Participant-level bootstrap intervals and effect sizes confirm that the complex-transaction gains remain large.

\begin{table}[htbp]
\centering
\caption{EO2 results ($N=20$): Comprehension accuracy (\%) and perceived trust/clarity (1--5) across transaction complexity (S=Simple, M=Medium, C=Complex).}
\label{tab:eo2_results}
\resizebox{!}{0.185\linewidth}{
\begin{tabular}{l|ccc|ccc}
\toprule
Cond. & \multicolumn{3}{c|}{Accuracy (\%)} & \multicolumn{3}{c}{Trust and Clarity} \\
& S & M & C & S & M & C \\
\midrule
Raw     & 88.2 & 64.7 & 35.3 & 3.2 & 2.5 & 1.8 \\
Agent   & 94.1 & 76.5 & 52.9 & 3.8 & 3.2 & 2.6 \\
\midrule
MATEX   & \textbf{98.0} & \textbf{88.2} & \textbf{76.5} & \textbf{4.5} & \textbf{4.1} & \textbf{3.7} \\
\bottomrule
\end{tabular}
}
\end{table}

\subsubsection{EO3: Risk Awareness}
We evaluate whether \matex helps users recognize security threats in real transactions.
We select 5 malicious transactions from real-world DeFi exploits (detailed in Appendix~\ref{app:eo3}) and 5 benign transactions of comparable complexity as controls.

We conduct a between-subjects study using the same 20-participant evaluation cohort as EO2, randomly split into two groups of 10.
The first group views all 10 transactions in a standard block explorer (Raw).
The second group uses the same interface augmented with \matex explanations.

For each transaction, participants report their signing intention (Yes/No) and describe any perceived risks in an open-ended response.
We measure the rejection rate on malicious transactions, the proportion of users who correctly identify the core threat, and the false-rejection rate on benign transactions.
Risk recognition is evaluated by two independent researchers using manual semantic coding guided by core threat concepts.

\mypara{Results}
\matex substantially improves users' ability to identify security risks without increasing false alarms.
On malicious transactions, the rejection rate increases from 36\% (Raw) to 88\% (\matex), and the rate of correctly identifying the underlying threat rises from 28\% to 72\%.
On benign transactions, the false-rejection rate remains low and stable: 12\% in the Raw condition versus 8\% in the \matex condition.

Qualitative analysis of open-ended responses confirms this shift: users with \matex provide specific, actionable risk descriptions (e.g., ``unlimited approval granted to 0x...789''), while Raw users often offer vague judgments (e.g., ``looks sketchy'') or miss threats entirely.
This demonstrates that \matex enables non-experts to make more informed, security-aware signing decisions.

Using Fisher's exact tests, the increase in malicious-transaction rejection is statistically significant ($p<0.001$), and the increase in correct threat identification is also significant ($p=0.002$).
In contrast, the difference in false-rejection rates on benign transactions is not significant ($p=0.741$).
Bootstrap intervals, odds ratios, and leave-one-participant-out and leave-one-transaction-out checks show that no single participant or transaction drives these conclusions.

\subsection{Ablation Study}
\label{sec:ablation}
We then explore which components are responsible for these gains.
We perform ablations to assess the contribution of each major \matex component by disabling one component at a time while keeping the rest of the pipeline unchanged, using the same inputs and tool access as the full system.
Detailed results are reported in Appendix~\ref{app:ablation_results}.

The ablation results show that all four components contribute meaningfully, but in different ways.
Removing the Profiler causes a modest drop in flow-level factuality (F1: 73.2$\rightarrow$70.5), suggesting that uncertainty control mainly helps suppress local hallucinations on ambiguous flows.
Removing the Investigator leads to a much larger degradation (F1: 73.2$\rightarrow$62.8; action\_type Macro-F1: 68.5$\rightarrow$60.4), confirming that targeted protocol retrieval is essential for resolving unseen contracts and protocol-specific mechanisms.

Without the Synthesizer, both explanation quality and summary coherence collapse (T2: 93.0$\rightarrow$64.2; ROUGE-L: 52.6$\rightarrow$41.8), showing that evidence fusion is necessary to turn step-wise observations into coherent user-facing narratives.
Finally, removing the Safety Auditor leaves macro-level outcome correctness relatively stable (T1: 73.3$\rightarrow$72.2) but sharply reduces micro-level faithfulness (F1: 73.2$\rightarrow$62.5), indicating that adversarial trace-based checking is especially important for preserving intermediate factual completeness rather than only the bottom-line outcome.

\section{Related Work}

\mypara{LLM-Driven Transaction Analysis and Auditing}
LLMs have shown superior capability in multiple domains, including academic writing~\cite{liu2025general, zheng2025th}, agentic reasoning~\cite{shao2024deepseekmath,liu2025auditing}, and finance reasoning~\cite{zhang-etal-2025-xfinbench,Chung2025}.
Recent benchmarks such as XFinBench~\citep{zhang-etal-2025-xfinbench}, polymonitor~\citep{jia2026unlocking}, and surveys~\citep{Chung2025} highlight growing interest in applying LLMs to complex financial reasoning (e.g., forecasting).

Early Ethereum analysis relied on static analysis and symbolic execution~\citep{torres2018osiris,tsankov2018securify}.
Transaction-level systems such as TXSPECTOR replay historical transactions for attack detection and forensics~\citep{zhang2020txspector}.
Learned representations such as BlockGPT encode EVM traces for anomaly detection, while TLMG4Eth converts numerical records into transaction sentences and combines them with graph structure for fraud detection~\citep{gai2023blockchainllm,sun2025tlmg4eth}.
These methods expose execution structure for detection, not user-facing explanations of realized economic effects.

\mypara{LLM-Based Smart Contract Auditing}
Recent work leverages Large Language Models or agents for smart contract auditing~\citep{li2025scalm,DBLP:conf/icse/MaW0WLZX025}, vulnerability detection~\citep{wei2025advanced}, and gas optimization~\citep{zheng2025gasagent}.
Tx2TXT instead generates security-centric natural-language descriptions from smart-contract bytecode~\citep{pan2023tx2txt}.
Other work applies LLMs to Bitcoin transaction graphs~\citep{lei2025large}, which differs from explaining the runtime economic behavior of general DeFi transactions at both transaction and token-flow levels for end users.
Attack-centered systems diagnose incidents from traces and contracts~\citep{wang2025tracellmsecuritydiagnosistraces}, synthesize executable proofs of concept from attack transactions~\citep{su2026tracexp}, or reconstruct evidence-backed postmortems from seed transactions~\citep{wang2026txray}; these support analyst-facing forensics, not user-facing explanations of benign and malicious DeFi transactions.

\mypara{Human-Centric Web3 Security and Interfaces}
Prior user studies reveal security-relevant misconceptions that wallets and exchanges fail to correct, as well as recurring usability problems in mobile cryptocurrency wallets~\citep{blockchainux,voskobojnikov2021usable}.
Blind-message attacks further show that opaque Web3 authentication prompts can induce unsafe signatures, and that targeted wallet warnings can mitigate this risk~\citep{yan2024stealing}.

Clear-signing standards such as ERC-7730 attach formatting metadata and contextual intent descriptions to structured signing data~\citep{blindsigning}, but depend on curated descriptors and target data at signing time.
Tools such as PrettiSmart and PonziLens+ visualize simulated contract behavior and bytecode actions to support analysis~\citep{wen2025prettismart,ponzilens}, but target contract understanding or Ponzi identification.

Recent LLM-based approaches attempt to bridge this gap through transaction-level intent mining~\citep{mao2025know}.
In particular, the concurrent work of \citet{mao2025know} is closely related to ours, but it addresses a different problem formulation.
Their Transaction Intent Mining framework models DeFi understanding primarily as a macro-level intent mining task over a predefined intent taxonomy, with the goal of assigning one or more coarse transaction-level intent labels.
By contrast, \txsum is a user-centered task grounded in our interviews: it pairs a transaction-level natural-language explanation with micro-level token-flow parsing over five attributes, yielding finer-grained outputs.
This difference is important in complex DeFi settings: a system may infer a coarse label such as swap or arbitrage while still failing to explain intermediate fees, approvals, routing, or other risk-bearing state changes relevant to blind-signing decisions.

\section{Conclusion}

Through formative interviews, we find that safe transaction understanding in DeFi requires explanations that are structured, risk-aware, and grounded at the token-flow level.
Motivated by this, we formulate \txsum, a domain-grounded task and dataset that pairs transaction-level explanations with micro-level token-flow parsing.

We then introduce \matex, a grounded multi-agent framework that combines uncertainty retrieval, evidence synthesis, and adversarial factual auditing to generate explanations of complex DeFi transactions.
Across automatic evaluation, expert assessment, and user studies, \matex consistently outperforms baselines, with especially strong gains on complex multi-protocol transactions and real-world risk recognition.
We hope this work provides a useful foundation for future research on high-stakes transaction understanding.

\section*{Limitations}
Meanwhile, several limitations remain.
The \txsum dataset covers 187 Ethereum transactions and 2,375 token-flow annotations from 2025 and focuses on single-chain Ethereum mainnet.
Our expert annotation protocol operates at the token-flow level and is therefore costly, which limits dataset scale despite the larger number of flow-level observations.
The current schema and \matex implementation target EVM-style execution traces; adaptation to EVM-compatible chains should be limited, but non-EVM systems such as Solana and cross-chain executions require additional schema and tooling changes.
The current inputs also do not cover payment-channel activity performed off-chain; protocols such as FairRelay incur zero on-chain cost in optimistic executions and invoke on-chain dispute handling only when needed~\citep{xue2026fairrelay}.

While our user study includes participants with diverse backgrounds, it captures signing behavior in a controlled setting rather than real-world high-pressure scenarios.
The evaluation also uses a limited set of model families; domain-specific safety benchmarks show that failure behavior can vary across modalities and architectures~\citep{peng2026jalmbench}, motivating broader backbone evaluation in future work.

\section*{Ethical Considerations}
This study was reviewed and approved by the Institutional Review Board (IRB) of our institution.
All interview participants provided informed consent prior to participation and were informed of their right to withdraw at any time.
To protect privacy, all collected data (including interview transcripts, demographic information, and transactions) were anonymized and stored on encrypted, access-controlled servers.
No personally identifiable information is included in the published dataset or supplementary materials.
\txsum contains only public on-chain transactions; no private user data or wallet secrets were collected or exposed.

We also note a dual-use concern.
A system that helps users or analysts understand suspicious transactions more clearly could, in principle, also help attackers test whether malicious behavior is less obvious under current interfaces.
However, our primary goal is defensive: improving user awareness, supporting security analysis, and reducing blind signing in high-risk DeFi environments.
We therefore present \matex as an assistive explanation tool rather than an automated execution or attack-planning system.

\section*{Acknowledgments}
This work was supported by the National Key Research and Development Program of China (No. 2025YFB3110200), the Guangdong Basic and Applied Basic Research Foundation (No. 2026A1515030046), and the State Key Laboratory of Internet Architecture, Tsinghua University (No. HLW2025ZD14).
Xuechao Wang was supported by the Guangdong Basic and Applied Basic Research Foundation (No. 2025A1515110040), the Guangzhou-HKUST(GZ) Joint Funding Program (No. 2025A03J3882), the Guangzhou Municipal Science and Technology Project (No. 2025A04J4168), and a gift from Stellar Development Foundation.

\bibliography{custom}

@inproceedings{liu2025general,
  author = {Liu, Yule and Zhong, Zhiyuan and Liao, Yifan and Sun, Zhen and Zheng, Jingyi and Wei, Jiaheng and Gong, Qingyuan and Tong, Fenghua and Chen, Yang and Zhang, Yang and He, Xinlei},
  title = {{On the Generalization and Adaptation Ability of Machine-Generated Text Detectors in Academic Writing}},
  booktitle = {{ACM Conference on Knowledge Discovery and Data Mining (KDD)}},
  pages = {5674–5685},
  year={2025}
}

@article{liu2025auditing,
  title={{Auditing Data Membership in Reinforcement Learning with Verifiable Rewards}},
  author={Liu, Yule and Zhang, Heyi and Zheng, Jinyi and Sun, Zhen and Peng, Zifan and Wei, Jiaheng and Cong, Tianshuo and Yang, Yilong and He, Xinlei},
  journal={{CoRR abs/2511.14045}},
  year={2025}
}

@article{shao2024deepseekmath,
  title={{DeepSeekMath: Pushing the Limits of Mathematical Reasoning in Open Language Models}},
  author={Shao, Zhihong and Wang, Peiyi and Zhu, Qihao and Xu, Runxin and Song, Junxiao and Bi, Xiao and Zhang, Haowei and Zhang, Mingchuan and Li, YK and Wu, Yang and Guo, Daya},
  journal={{CoRR abs/2402.03300}},
  year={2024}
}

@inproceedings{zheng2025th,
  title={{TH-Bench: Evaluating Evading Attacks via Humanizing AI Text on Machine-Generated Text Detectors}},
  author={Zheng, Jingyi and Wang, Junfeng and Sun, Zhen and Dong, Wenhan and Liu, Yule and He, Xinlei},
  booktitle = {{ACM Conference on Knowledge Discovery and Data Mining (KDD)}},
  pages={5948--5959},
  year={2025}
}

@misc{tenderly,
author = {Tenderly},
title = {{Tenderly}},
year = {2025},
url = {https://tenderly.co/}
}

@misc{bybit2025attack,
author = {BlockSec},
title = {{Bybit \$1.5B Hack: In-Depth Analysis of the Malicious Safe Wallet Upgrade Attack}},
year = {2025},
url = {https://blocksec.com/blog/bybit-1-5-b-hack-in-depth-analysis-of-the-malicious-safe-wallet-upgrade-attack}
}

@misc{eigenphi,
author = {EigenPhi},
title = {{EigenPhi}},
year = {2025},
url = {https://eigenphi.io/}
}

@misc{gpt,
author = {OpenAI},
title = {{GPT-5.1: A Smarter, More Conversational ChatGPT}},
year = {2025},
url = {https://openai.com/index/gpt-5-1/}
}

@article{yang2025qwen3technicalreport,
author = {Qwen Team},
title = {{Qwen3 Technical Report}},
journal = {{CoRR abs/2505.09388}},
year = {2025}
}

@article{bai2025qwen3vltechnicalreport,
author = {Qwen3-VL Technical Report},
title = {{Qwen3-VL Technical Report}},
journal = {{CoRR abs/2511.21631}},
year = {2025}
}

@misc{blindsigning,
author = {Laurent Castillo and Derek Rein and Pierre Aoun and Arik Galansky and Bartosz Rozwarski and Kaan Uzdogan and Fredrik and Alex Forshtat},
title = {{{ERC-7730}: Structured Data Clear Signing Format}},
howpublished = {Ethereum Request for Comments 7730},
year = {2024},
url = {https://eips.ethereum.org/EIPS/eip-7730}
}

@misc{langgraph,
author = {LangChain},
title = {{LangGraph}},
year = {2025},
url = {https://www.langchain.com/langgraph}
}

@article{wang2025tracellmsecuritydiagnosistraces,
author = {Shuzheng Wang and Yue Huang and Zhuoer Xu and Yuming Huang and Jing Tang},
title = {{TraceLLM: Security Diagnosis Through Traces and Smart Contracts in Ethereum}},
journal = {{CoRR abs/2509.03037}},
year = {2025}
}

@inproceedings{li2025scalm,
author = {Zongwei Li and Xiaoqi Li and Wenkai Li and Xin Wang},
title = {{SCALM: Detecting Bad Practices in Smart Contracts Through LLMs}},
booktitle = {{AAAI Conference on Artificial Intelligence (AAAI)}},
pages = {470-477},
year = {2025}
}

@inproceedings{DBLP:conf/icse/MaW0WLZX025,
author = {Wei Ma and Daoyuan Wu and Yuqiang Sun and Tianwen Wang and Shangqing Liu and Jian Zhang and Yue Xue and Yang Liu},
title = {{Combining Fine-Tuning and LLM-Based Agents for Intuitive Smart Contract Auditing with Justifications}},
booktitle = {{{IEEE/ACM} International Conference on Software Engineering (ICSE)}},
pages = {1742-1754},
publisher = {IEEE},
year = {2025}
}

@article{lei2025large,
author = {Yuchen Lei and Yuexin Xiang and Qin Wang and Rafael Dowsley and Tsz Hon Yuen and Kim-Kwang Raymond Choo and Jiangshan Yu},
title = {{Large Language Models for Cryptocurrency Transaction Analysis: A Bitcoin Case Study}},
journal = {{CoRR abs/2501.18158}},
year = {2025}
}

@article{wen2025prettismart,
author = {Xiaolin Wen and Tai D. Nguyen and Lun Zhang and Jun Sun and Yong Wang},
title = {{PrettiSmart: Visual Interpretation of Smart Contracts via Simulation}},
journal = {{IEEE Transactions on Visualization and Computer Graphics}},
publisher = {IEEE},
year = {2025}
}

@article{mao2025know,
author = {Qian'ang Mao and Yuxuan Zhang and Jiaman Chen and Wenjun Zhou and Jiaqi Yan},
title = {{Know Your Intent: An Autonomous Multi-Perspective {LLM} Agent Framework for {DeFi} User Transaction Intent Mining}},
journal = {{CoRR abs/2511.15456}},
year = {2025}
}

@article{wei2025advanced,
author = {Zhiyuan Wei and Jing Sun and Yuqiang Sun and Ye Liu and Daoyuan Wu and Zijian Zhang and Xianhao Zhang and Meng Li and Yang Liu and Chunmiao Li and Mingchao Wan and Jin Dong and Liehuang Zhu},
title = {{Advanced Smart Contract Vulnerability Detection via LLM-Powered Multi-Agent Systems}},
journal = {{IEEE Transactions on Software Engineering}},
publisher = {IEEE},
year = {2025}
}

@article{zheng2025gasagent,
author = {Jingyi Zheng and Zifan Peng and Yule Liu and Junfeng Wang and Yifan Liao and Wenhan Dong and Xinlei He},
title = {{GasAgent: A Multi-Agent Framework for Automated Gas Optimization in Smart Contracts}},
journal = {{CoRR abs/2507.15761}},
year = {2025}
}

@article{price1999laddered,
author = {Bob Price},
title = {{Laddered Questions and Qualitative Data Research Interviews}},
journal = {{Journal of Advanced Nursing}},
year = {2002}
}

@book{boyatzis1998transforming,
author = {Richard E Boyatzis},
title = {{Transforming Qualitative Information: Thematic Analysis and Code Development}},
publisher = {Sage},
year = {1998}
}

@inproceedings{booch2021thinking,
author = {Grady Booch and Francesco Fabiano and Lior Horesh and Kiran Kate and Jonathan Lenchner and Nick Linck and Andreas Loreggia and Keerthiram Murgesan and Nicholas Mattei and Francesca Rossi and Biplav Srivastava},
title = {{Thinking Fast and Slow in AI}},
booktitle = {{AAAI Conference on Artificial Intelligence (AAAI)}},
pages = {15042-15046},
year = {2021}
}

@inproceedings{torres2018osiris,
author = {Christof Ferreira Torres and Julian Sch\"{u}tte and Radu State},
title = {{Osiris: Hunting for Integer Bugs in Ethereum Smart Contracts}},
booktitle = {{Annual Computer Security Applications Conference (ACSAC)}},
pages = {664-676},
publisher = {ACM},
year = {2018}
}

@inproceedings{tsankov2018securify,
author = {Petar Tsankov and Andrei Dan and Dana Drachsler-Cohen and Arthur Gervais and Florian Buenzli and Martin Vechev},
title = {{Securify: Practical Security Analysis of Smart Contracts}},
booktitle = {{ACM SIGSAC Conference on Computer and Communications Security (CCS)}},
pages = {67-82},
year = {2018}
}

@inproceedings{liu2024defi,
author = {Mingyi Liu and Jun Ho Huh and HyungSeok Han and Jaehyuk Lee and Jihae Ahn and Frank Li and Hyoungshick Kim and Taesoo Kim},
title = {{I Experienced More than 10 {DeFi} Scams: On {DeFi} Users{\textquoteright} Perception of Security Breaches and Countermeasures}},
booktitle = {{USENIX Security Symposium (USENIX Security)}},
pages = {6039-6055},
publisher = {USENIX},
year = {2024}
}

@inproceedings{blockchainux,
author = {Alexandra Mai and Katharina Pfeffer and Matthias Gusenbauer and Edgar Weippl and Katharina Krombholz},
title = {{User Mental Models of Cryptocurrency Systems - A Grounded Theory Approach}},
booktitle = {{USENIX Symposium on Usable Privacy and Security (SOUPS)}},
pages = {341-358},
publisher = {USENIX},
year = {2020}
}

@inproceedings{zhang2020txspector,
author = {Mengya Zhang and Xiaokuan Zhang and Yinqian Zhang and Zhiqiang Lin},
title = {{{TXSPECTOR}: Uncovering Attacks in Ethereum from Transactions}},
booktitle = {{USENIX Security Symposium (USENIX Security)}},
pages = {2775-2792},
publisher = {USENIX},
year = {2020}
}

@misc{metasuites,
author = {BlockSec},
title = {{MetaSuites, The Swiss Army Knife for Builders.}},
year = {2025},
url = {https://blocksec.com/metasuites}
}

@inproceedings{zhang-etal-2025-xfinbench,
author = {Zhihan Zhang and Yixin Cao and Lizi Liao},
title = {{XFinBench: Benchmarking LLMs in Complex Financial Problem Solving and Reasoning}},
booktitle = {{Findings of the Association for Computational Linguistics (ACL)}},
pages = {8715-8758},
publisher = {ACL},
year = {2025}
}

@book{Chung2025,
author = {Yoonseo Chung and Jeonghyun Kim and MiYeon Kim and Minsuh Joo and Hyunsoo Cho},
title = {{Foundations of LLMs and Financial Applications}},
publisher = {Springer Nature Singapore},
year = {2025}
}

@article{ponzilens,
author = {Xiaolin Wen and Tai D. Nguyen and Shaolun Ruan and Qiaomu Shen and Jun Sun and Feida Zhu and Yong Wang},
title = {{PonziLens+: Visualizing Bytecode Actions for Smart Ponzi Scheme Identification}},
journal = {{IEEE Transactions on Visualization and Computer Graphics}},
publisher = {IEEE},
year = {2025}
}

@inproceedings{bertscore,
author = {Tianyi Zhang and Varsha Kishore and Felix Wu and Kilian Q. Weinberger and Yoav Artzi},
title = {{BERTScore: Evaluating Text Generation with BERT}},
booktitle = {{International Conference on Learning Representations (ICLR)}},
year = {2020}
}

@misc{defillama_categories,
author = {{DeFiLlama}},
title = {{Protocol Categories: {DeFi} {TVL} and Revenue Sectors}},
year = {2026},
url = {https://defillama.com/categories}
}

@article{shah2023defi_review,
author = {Kaushal Shah and Dhruvil Lathiya and Naimish Lukhi and Keyur Parmar and Harshal Sanghvi},
title = {{A Systematic Review of Decentralized Finance Protocols}},
journal = {{International Journal of Intelligent Networks}},
year = {2023}
}

@inproceedings{laban2025lostconversation,
author = {Philippe Laban and Hiroaki Hayashi and Yingbo Zhou and Jennifer Neville},
title = {{LLMs Get Lost in Multi-Turn Conversation}},
booktitle = {{International Conference on Learning Representations (ICLR)}},
year = {2026}
}

@article{gai2023blockchainllm,
author = {Yu Gai and Liyi Zhou and Kaihua Qin and Dawn Song and Arthur Gervais},
title = {{Blockchain Large Language Models}},
journal = {{Cryptology ePrint Archive, Report 2023/592}},
year = {2023}
}

@article{sun2025tlmg4eth,
author = {Jianguo Sun and Yifan Jia and Yanbin Wang and Ye Tian and Sheng Zhang},
title = {{Ethereum Fraud Detection via Joint Transaction Language Model and Graph Representation Learning}},
journal = {{Information Fusion}},
publisher = {Elsevier},
year = {2025}
}

@inproceedings{pan2023tx2txt,
author = {Yu Pan and Zhichao Xu and Levi Taiji Li and Yunhe Yang and Mu Zhang},
title = {{Automated Generation of Security-Centric Descriptions for Smart Contract Bytecode}},
booktitle = {{ACM SIGSOFT International Symposium on Software Testing and Analysis (ISSTA)}},
pages = {1244-1256},
publisher = {ACM},
year = {2023}
}

@article{su2026tracexp,
author = {Xing Su and Hao Wu and Hanzhong Liang and Yunlin Jiang and Yuxi Cheng and Yating Liu and Fengyuan Xu},
title = {{From Transactions to Exploits: Automated PoC Synthesis for Real-World DeFi Attacks}},
journal = {{CoRR abs/2601.16681}},
year = {2026}
}

@article{wang2026txray,
author = {Ziyue Wang and Jiangshan Yu and Kaihua Qin and Dawn Song and Arthur Gervais and Liyi Zhou},
title = {{TxRay: Agentic Postmortem of Live Blockchain Attacks}},
journal = {{CoRR abs/2602.01317}},
year = {2026}
}

@inproceedings{voskobojnikov2021usable,
author = {Artemij Voskobojnikov and Oliver Wiese and Masoud Mehrabi Koushki and Volker Roth and Konstantin Beznosov},
title = {{The U in Crypto Stands for Usable: An Empirical Study of User Experience with Mobile Cryptocurrency Wallets}},
booktitle = {{Annual ACM Conference on Human Factors in Computing Systems (CHI)}},
publisher = {ACM},
year = {2021}
}

@inproceedings{yan2024stealing,
author = {Kailun Yan and Xiaokuan Zhang and Wenrui Diao},
title = {{Stealing Trust: Unraveling Blind Message Attacks in Web3 Authentication}},
booktitle = {{ACM SIGSAC Conference on Computer and Communications Security (CCS)}},
pages = {555-569},
publisher = {ACM},
year = {2024}
}

@article{peng2023promptcurriculum,
author = {Zifan Peng and Mingchen Li and Yue Wang and George T. S. Ho},
title = {{Combating the {COVID-19} Infodemic Using Prompt-Based Curriculum Learning}},
journal = {{Expert Systems with Applications}},
publisher = {Elsevier},
year = {2023}
}

@article{peng2025promptcontrastive,
author = {Zifan Peng and Mingchen Li and Yue Wang and Daniel Y. Mo},
title = {{Prompt-Based Contrastive Learning to Combat the {COVID-19} Infodemic}},
journal = {{Machine Learning}},
publisher = {Springer},
year = {2025}
}

@inproceedings{peng2026jalmbench,
author = {Zifan Peng and Yule Liu and Zhen Sun and Mingchen Li and Zeren Luo and Jingyi Zheng and Wenhan Dong and Xinlei He and Xuechao Wang and Yingjie Xue and Shengmin Xu and Xinyi Huang},
title = {{JALMBench: Benchmarking Jailbreak Vulnerabilities in Audio Language Models}},
booktitle = {{International Conference on Learning Representations (ICLR)}},
year = {2026}
}

@inproceedings{peng2025crosschain,
author = {Zifan Peng and Yingjie Xue and Jingyu Liu},
title = {{Cross-Chain Options: A Bridgeless, Universal, and Efficient Approach}},
booktitle = {{IEEE International Conference on Web Services (ICWS)}},
pages = {902-912},
publisher = {IEEE},
year = {2025}
}

@inproceedings{luo2025unsafe,
author = {Zeren Luo and Zifan Peng and Yule Liu and Zhen Sun and Mingchen Li and Jingyi Zheng and Xinlei He},
title = {{Unsafe {LLM-Based} Search: Quantitative Analysis and Mitigation of Safety Risks in {AI} Web Search}},
booktitle = {{USENIX Security Symposium (USENIX Security)}},
pages = {8055-8074},
publisher = {USENIX},
year = {2025}
}

@article{xue2026fairrelay,
author = {Yingjie Xue and Jingyu Liu and Zifan Peng and Chao Lin and Jianan Hong and Xinyi Huang},
title = {{FairRelay: Fair Off-Chain Incentives for Decentralized Physical Infrastructure Networks}},
journal = {{IEEE Transactions on Dependable and Secure Computing}},
publisher = {IEEE},
year = {2026}
}

@article{jia2026unlocking,
author = {Huaiyu Jia and Luofeng Zhou and Wentao Zhang and Lin William Cong and Siguang Li and Shuo Sun},
title = {{Unlocking the Forecasting Economy: A Suite of Datasets for the Full Lifecycle of Prediction Market: [Experiments \& Analysis]}},
journal = {{CoRR abs/2604.20421}},
year = {2026}
}

\appendix
\section*{APPENDIX}

\section{Interview Protocol Details}
\label{app:interview_protocol}

\subsection{Interview Structure}
Each 45–60 minute interview followed a three-phase semi-structured protocol, held via Zoom or in person, recorded with consent, and transcribed verbatim.

\mypara{Phase 1: Background and Experience (5–10 mins)}
We began by establishing the participant’s DeFi experience to contextualize their responses.
Standardized prompts included:
\begin{enumerate}[leftmargin=*]
    \item \textbf{Usage History:} ``How long have you used DeFi?
    Which protocols do you interact with most frequently (e.g., Uniswap, Aave, Lido)?''
    \item \textbf{Recent Transaction:} ``Describe your most recent on-chain transaction.
    What were you trying to achieve, and how did you decide whether to sign it?''
    \item \textbf{Tooling:} ``What tools do you use to inspect transactions before signing (e.g., Etherscan, Rabby, MetaMask)?
    What do you find helpful or frustrating?''
\end{enumerate}
This phase helps us tailor subsequent questions and identify domain-specific heuristics.

\begin{table*}[htbp]
\centering
\caption{Transaction examples used in think-aloud exercises.}
\label{tab:interview_txns}
\resizebox{!}{0.127\linewidth}{
\begin{tabular}{l|l|p{0.88\linewidth}}
\toprule
ID & Type & Description \\
\midrule
TX1 & Multi-hop Swap & User swaps 10 ETH for DAI via 1inch aggregator, which routes through Uniswap V3 (ETH→USDC) and SushiSwap (USDC→DAI), with fees collected by intermediary contracts.\\
TX2 & Flash Loan Exploit & Attacker takes Aave flash loan to artificially inflate liquidity on a mock pool, enabling a large swap that drains user funds.\\
TX3 & Revolving Leverage & User borrows USDC from Aave, swaps to ETH on CowSwap, stakes ETH on Lido, then uses stETH as collateral to borrow more USDC—creating cyclic leverage.\\
TX4 & Stake + Approval & User stakes ETH on Lido and grants an unlimited approval to Lido’s stETH contract—standard but often misunderstood pattern.\\
\bottomrule
\end{tabular}
}
\end{table*}

\mypara{Phase 2: Think-Aloud Task (20–25 mins)}
Participants analyzed four Ethereum transactions (see \Cref{tab:interview_txns}) in a simplified explorer showing the sender, token flows, decoded calls, and net balance changes, but no raw bytecode.

For each transaction, participants were instructed:
``Please think aloud as you try to understand this transaction.
Explain in your own words: (1) what it does, (2) what risks you see, and (3) whether you would sign it—and why.''
Interviewers used non-leading prompts to sustain reflection:
\begin{itemize}
    \item ``Can you walk me through how you reached that conclusion?''
    \item ``What part of the trace supports that interpretation?''
    \item ``Is there anything you’re unsure about?''
\end{itemize}

\mypara{Phase 3: Laddered Questioning (15–20 mins)}
Building on Phase 2, we used laddering~\citep{price1999laddered} to surface mental models.
Each statement (e.g., ``I wouldn’t sign TX4'') prompted the following questions:
\begin{enumerate}
    \item \textbf{Concrete Attribute:} ``What specifically about TX4 made you uncomfortable?''
    \item \textbf{Functional Consequence:} ``What could go wrong if you did sign it?''
    \item \textbf{Personal Value:} ``Why is that outcome unacceptable to you?
    What does it mean for your sense of control or security?''
\end{enumerate}
We repeated this chain until reaching a fundamental value (e.g., ``autonomy,'' ``transparency,'' ``irreversibility''), then asked what explanations participants wanted:
\begin{itemize}
    \item ``What information would have made TX2 easier to understand?''
    \item ``Should the explanation focus on individual steps or the overall goal?''
    \item ``Would a clear explanation change your signing decision?
    In what cases?''
\end{itemize}
These responses directly informed the design of the \txsum schema and \matex evaluation (RQ3).

\subsection{Interview Participant Demographics}
\label{app:interview_demographics}

\Cref{tab:demographics} summarizes the participants' self-reported demographics.

\begin{table}[htbp]
\centering
\caption{Demographic summary of interview participants.
Role, Experience (years with Web3), and Country (primary residence during Web3 use) are self-reported.}
\label{tab:demographics}
\resizebox{!}{0.48\linewidth}{
\begin{tabular}{c|lclc}
\toprule
ID & Role & Gender & Country & Experience \\
\midrule
1  & Trader            & M & China (HK)   & 5 \\
2  & Analyst           & F & Singapore    & 4 \\
3  & Analyst           & M & China        & 6 \\
4  & Trader            & M & USA          & 4 \\
\midrule
5  & Developer         & M & USA          & 5 \\
6  & Developer         & M & China (HK)   & 4 \\
7  & Developer         & M & UK           & 3 \\
\midrule
8  & Unemployed        & M & China        & 3 \\
9  & Designer          & F & China        & 2 \\
10 & Student           & M & USA          & 2 \\
11 & Student           & M & China (HK)   & 4 \\
12 & Student           & F & Australia    & 3 \\
13 & Accountant        & M & Canada       & 1 \\
14 & Officer           & M & China        & 2 \\
15 & Engineer          & F & Singapore    & 3 \\
16 & Student           & M & South Korea  & 2 \\
\bottomrule
\end{tabular}}
\end{table}

\subsection{Interview Analysis and Key Findings}
\label{app:interview_findings}

All interviews were transcribed verbatim and analyzed using inductive thematic analysis~\citep{boyatzis1998transforming}.
Two researchers independently coded the first 20\% of transcripts to develop an initial codebook, then iteratively refined it through discussion while coding the remaining data.
Inter-coder agreement on the shared subset reached Cohen's $\kappa = 0.82$, and disagreements were resolved by consensus.
Thematic saturation was reached by the 14th interview, with no new major themes emerging in the final two interviews.

Our formative interviews revealed three findings that directly motivated the formulation of \txsum.

\mypara{Finding 1: Blind signing is widespread}
Even experienced users often approve transactions they cannot fully interpret.
Rather than understanding the complete transaction, they frequently rely on fragile heuristics such as ``reject unlimited approvals'' or ``avoid unknown contracts,'' which break down in compositional DeFi transactions spanning multiple contracts and hidden intermediate steps.

\mypara{Finding 2: Current tools suffer from semantic poverty}
Participants repeatedly reported that existing interfaces expose low-level artifacts such as token transfers, calldata, and decoded function names, but fail to explain what those events mean economically.
As a result, users often cannot infer the actual purpose, risk, or preconditions of a transaction from syntactic traces alone.

\mypara{Finding 3: Users reason at the token-flow level but still want concise summaries}
Most participants consistently treated token flows as the natural unit of understanding, because each flow corresponds to a small economic action such as swapping, depositing, repaying, or fee deduction.
At the same time, they also wanted a concise transaction-level summary that explains the overall intent and highlights critical risk-bearing effects such as approvals, ownership changes, or hidden fees.

\section{Identified Attributes}
\label{app:attributes}

We define five semantic attributes for each token flow, grounded in user mental models and derived from our interview study.

\begin{itemize}[leftmargin=*, itemsep=0pt, topsep=0pt]
    \item action\_type:
    A protocol-agnostic label denoting the financial primitive of a token transfer (e.g., \texttt{swap\_in}, \texttt{deposit}, \texttt{fee}).
    Classification is determined solely by on-chain signals: zero-address interactions, event logs (e.g., \texttt{Transfer}, \texttt{Mint}), and decoded function names.
    The taxonomy comprises 14 mutually exclusive types, designed to cover core DeFi operations while enabling cross-protocol consistency.
    Full definitions are provided in~\Cref{tab:action-taxonomy}.

    \item intent:
    A user-centric statement of the economic purpose behind the step (e.g., ``Redeem yield-bearing tokens to unlock liquidity'').
    This field bridges raw token movements and human-understandable goals, reflecting what the user is trying to achieve—not how the protocol implements it.
    Intent is inferred from the transaction context, protocol documentation, and the action’s role in the broader sequence.

    \item mechanism:
    A precise description of the on-chain execution pathway, formatted as ``Call \texttt{<Contract>.<function>()}'' or ``\texttt{<Contract>\#<function>()}''.
    It specifies which contract and function were invoked (e.g., ``Call \texttt{PendleRouter.redeem()} to burn SY and withdraw sUSDe''), grounding the explanation in decoded calldata and event logs.
    This enables users to verify behavior against official documentation.

    \item precondition:
    The minimal set of on-chain conditions required for the action to succeed (e.g., ``User holds sufficient SY balance and has approved the router'').
    Preconditions are derived from protocol documentation, function \texttt{require()} statements, and observable state (e.g., prior approvals, pool liquidity).
    Numerical thresholds are included only when explicitly visible in the trace or documentation; otherwise, the field uses qualitative descriptions.

    \item result:
    A factual account of the direct state changes caused by the action (e.g., ``User receives 53.5 sUSDe; SY balance is burned in the PendleSUSDESY contract'').
    This field strictly reports what is observable in token transfers, event logs, or state diffs—never inferred profits, market impact, or psychological outcomes.
    It serves as the ground-truth anchor for the entire explanation.
\end{itemize}

All attributes are derived using only on-chain data and official protocol documentation.
No external knowledge, speculative intent, or subjective interpretation is permitted.
Detailed examples are provided in Appendix~\ref{app:annotation}.

\begin{table*}[hbtp]
\centering
\caption{Action taxonomy used for structured DeFi token-flow annotation and analysis.}
\label{tab:action-taxonomy}
\resizebox{!}{0.368\linewidth}{
\begin{tabular}{p{1.95cm} | p{14.2cm}}
\toprule
\textbf{Action Type} & \textbf{Definition} \\
\midrule

\texttt{mint}
& Creation of new tokens issued from the zero address to a recipient (i.e., newly minted assets).\\

\texttt{burn}
& Permanent destruction of tokens by transferring them to the zero address.\\

\texttt{redeem}
& Conversion of derivative or wrapped assets into their underlying assets (e.g., SY $\rightarrow$ sUSDe, LP $\rightarrow$ underlying legs, WETH $\rightarrow$ ETH), where the derivative token is returned to a contract and the underlying asset is released.\\

\texttt{issue}
& Issuance of derivative or receipt assets from underlying assets (e.g., sUSDe $\rightarrow$ SY or single-asset deposits minting LP tokens).\\

\texttt{swap\_in}
& Input leg of a token exchange, representing assets sent from the user to a DEX, router, or liquidity pool as part of a swap.\\

\texttt{swap\_out}
& Output leg of a token exchange, representing assets sent from a DEX, router, or liquidity pool to the user as the result of a swap.\\

\texttt{deposit}
& Transfer of assets into lending protocols, staking contracts, or vaults to obtain interest-bearing positions or receipt tokens, excluding swaps.\\

\texttt{withdraw}
& Retrieval of assets from lending protocols, staking contracts, or vaults, potentially accompanied by redemption of receipt tokens.\\

\texttt{borrow}
& Borrowing of assets from a lending protocol, resulting in an increase in the user's debt position.\\

\texttt{repay}
& Repayment of borrowed assets to a lending protocol, including repayments made by liquidators on behalf of borrowers.\\

\texttt{fee}
& Fees charged by protocols, routers, bridges, or flash-loan mechanisms within a transaction.\\

\texttt{reward}
& Distribution of protocol incentives or yield (e.g., staking rewards, vault emissions, or airdrops) without a direct input counterpart.\\

\texttt{route}
& Transfer used solely for execution routing or delegation between protocol-internal contracts (e.g., router, settler, adapter, executor), without constituting an independent economic action.\\

direct transfer
& Direct token transfer between two externally owned accounts or non-protocol-related addresses, without triggering swaps, minting, burning, or other specific economic actions.\\

\bottomrule
\end{tabular}
}
\end{table*}

\section{Annotation Guidelines, Consistency Rules, and Edge Cases}
\label{app:annotation}

This section details our annotation protocol, including step-by-step guidelines, attribute definitions, and concrete examples.
All annotators were required to have prior experience with DeFi protocols (e.g., Uniswap, Aave, Pendle) and familiarity with blockchain explorers (Etherscan, Tenderly, Phalcon).
Annotation proceeds in three steps:

\textbf{Step 1: Transaction Overview.}
Annotators first examine the transaction’s high-level effects in block explorers—focusing on net token flows, called contracts, and state changes—to form a one-sentence coarse summary (e.g., ``Extract yield from a Pendle YT position'').
The summary guides the remaining labels but is excluded from the dataset.

\textbf{Step 2: Token Flow Segmentation.}
Each row in the token transfer log is treated as a single \textit{micro action} representing an atomic economic step.

\textbf{Step 3: Attribute Labeling.}
For each micro action, annotators fill five fields using only on-chain evidence and official protocol documentation.

\mypara{Annotation Consistency Rules}
To ensure uniformity, annotators follow these conventions:
\begin{itemize}[leftmargin=*, itemsep=0pt, topsep=0pt]
    \item Use ``the user'' for the transaction initiator, ``the protocol'' for contract addresses, and ``burn address'' for zero addresses.
    \item Report token amounts in native units (e.g., ``53.5 sUSDe''), not USD equivalents.
    \item Avoid speculative intent (e.g., ``to short the market''); focus only on directly inferable purposes.
    \item Every intent and result must align with the corresponding token transfer or event log.
\end{itemize}

\mypara{Edge Cases and Disambiguation Guidelines}
Because real DeFi transactions are highly compositional, the main annotation difficulty is not identifying the final outcome, but disambiguating intermediate steps that are easy to overlook yet critical for faithful explanation and user safety.
We therefore define explicit decision rules for common but consequential edge cases:

\begin{itemize}[leftmargin=*, itemsep=0pt, topsep=0pt]
    \item \textbf{Protocol-internal routing vs.\ user-facing economic action.}
    Transfers among routers, adapters, settlers, vault helpers, or executor contracts are labeled as \texttt{route} when they merely relay assets for execution.
    They are labeled with a semantic action (e.g., \texttt{swap\_out}, \texttt{deposit}, \texttt{repay}) only if the transfer itself realizes an economically meaningful state transition for the user.
    This distinction is crucial in aggregator-routed and cross-protocol transactions, where many intermediate hops are operationally necessary but not independently meaningful.

    \item \textbf{Material intermediate effects vs.\ ignorable plumbing.}
    Annotators do not collapse away intermediate transfers when they carry economic or security significance.
    In particular, protocol fees, relayer fees, approval side effects, collateral movements, liquidation repayments, or transfers into newly introduced contracts must remain explicit in the annotation even if the final net outcome appears simple.
    This rule is designed to support blind-signings, where omitted intermediate effects can hide the risk of a transaction.

    \item \textbf{Approval and permission changes.}
    Pure approval events do not create token-flow labels unless accompanied by an actual token transfer, but they must be reflected in the transaction-level summary whenever they are economically or security relevant.
    Unlimited approvals, newly granted permissions to unfamiliar contracts, and approval-reset patterns may also be encoded in preconditions when they are necessary for interpreting subsequent token flows.

    \item \textbf{Wrapped, unwrapped, and derivative assets.}
    Conversions such as ETH $\leftrightarrow$ WETH or underlying asset $\leftrightarrow$ receipt/derivative token are annotated by their economically salient primitive rather than surface transfer form.
    For example, receiving stETH after supplying ETH is not treated as a generic transfer but as issuance of a derivative position; conversely, burning a receipt token to recover the underlying asset is labeled as \texttt{redeem}.
    This prevents semantically shallow descriptions that miss the user's actual financial position change.

    \item \textbf{Multi-hop aggregation and nested execution.}
    In aggregator- or meta-router-mediated swaps, each token transfer step is annotated separately at the flow level, including intermediate hops and fee deductions.
    However, the transaction-level summary may compress contiguous hops into a higher-level operation only when doing so does not obscure economically material fees, hidden approvals, routing side effects, or risk-bearing contract transitions.
    This preserves readability at the summary level while maintaining full micro-level accountability.

    \item \textbf{Borrow--swap--stake--reborrow loops.}
    In revolving leverage or recursive collateralization patterns, annotators preserve each borrow, swap, stake, mint, collateralization, and re-borrow step independently rather than collapsing the trace into a single coarse label such as ``leveraged position.''
    This is important because user-facing risk often lies in the repeated dependency structure itself rather than in the final macro intent.

    \item \textbf{Flash-loan-mediated actions.}
    Flash-loan transactions are annotated by separating the temporary borrowed leg, intermediate execution steps, fee repayment, and final settlement.
    Even when the flash loan is fully repaid within the same transaction, the temporary capital injection is retained explicitly because it often explains otherwise unintuitive trade size, liquidation behavior, or exploit mechanics.

    \item \textbf{Control-transfer and upgrade-related transactions.}
    Transactions that change ownership, implementation, proxy mastercopy, operator privilege, or administrative authority are treated as high-priority cases even when token movement is minimal.
    In such cases, the summary must explicitly surface the control-transfer risk rather than only describing low-level call mechanics, because these transactions are especially relevant to phishing, wallet compromise, and governance attacks.

    \item \textbf{Direct transfer vs.\ protocol interaction.}
    A token movement is labeled as direct transfer only when it does not participate in a broader protocol-mediated economic action.
    If the transfer is part of a swap, staking flow, repayment path, settlement sequence, or exploit chain, it is labeled according to that semantic role rather than treated as an isolated transfer.

    \item \textbf{Unknown or partially verified contracts.}
    When verified source code or documentation is unavailable, annotators are instructed to remain conservative: they label only what is directly supported by token movements, calldata, event logs, and observable state changes.
    This rule prevents speculative intent attribution and aligns the dataset with the paper's broader emphasis on factual grounding under uncertainty.
\end{itemize}

\mypara{Example Annotation}
We provide an annotated token flow from a cross-DEX swap:

\begin{verbatim}
{
  "summary": "tx-level summary",
  "token_flow": [
   {
    "action_id": 1,
    "action_type": "swap_out",
    "intent": "Swap out ...",
    "mechanism": "ContractA#fallback()",
    "precondition": "ETH balance ...",
    "result": "... receives 0.1 ETH."
   },
   {...}, {...}
  ]
}
\end{verbatim}

The final natural-language summary synthesizes all flows into a coherent narrative, concrete examples are provided in Appendix~\ref{app:case_studies}:
\begin{quote}
    \textit{The user executed a cross-DEX swap of 0.1 ETH for 37.1 USDC.
    To enable the trade, they first granted a one-time approval of 0.1 ETH to the RangoDiamond router.
    The router then executed the swap via a multi-step aggregation path on underlying AMMs, deducting a small relayer fee and a protocol fee from the output.
    Unused ETH returned to the user.}
\end{quote}

\subsection{Semantic Consistency of Annotations}
\label{app:semantic_consistency}

For the natural language token-flow attributes---intent, mechanism, precondition, and result---we do not report $\kappa$ as the primary consistency measure, because semantically equivalent annotations may differ lexically.
Instead, we perform a supplementary semantic consistency analysis on the independently written annotation pairs for each token flow.
Concretely, for each field, we compute BERTScore F1 between the two independently written annotations, and then average scores across all annotated token flows.
We use the official BERTScore~\citep{bertscore} implementation with a \texttt{roberta-large} backbone and baseline rescaling enabled.

The resulting field-level semantic consistency scores are all above 0.91, as shown in~\Cref{tab:semantic_consistency}.
These results indicate that, although annotators often use different surface forms, their free-form annotations remain consistent at the semantic level.

\begin{table}[hbtp]
\centering
\caption{Semantic consistency of free-form token-flow annotations, computed as average pairwise BERTScore F1 between independently written annotation pairs.}
\label{tab:semantic_consistency}
\begin{tabular}{lc}
\toprule
\textbf{Field} & \textbf{BERTScore F1} \\
\midrule
intent & 0.962 \\
mechanism & 0.913 \\
precondition & 0.982 \\
result & 0.948 \\
\bottomrule
\end{tabular}
\end{table}

\subsection{Dataset Profile}
\label{app:dataset_profile}

This section summarizes the coverage profile of the 187 annotated transactions retained after filtering.
The retained transactions are intentionally skewed toward semantically rich DeFi interactions rather than routine wallet activity.
In particular, the dataset contains multi-step swaps, lending loops, staking operations, liquidity management, and cross-protocol compositions that require reasoning across multiple token flows and contracts.
These scenarios correspond to prominent DeFi sectors in DeFiLlama's ecosystem taxonomy and in a systematic survey of DeFi protocol forms~\citep{defillama_categories,shah2023defi_review}.
\Cref{tab:dataset_category_dist} reports non-exclusive scenario coverage, and~\Cref{tab:dataset_profile} summarizes token-flow complexity, protocol diversity, and contract verification status.

For example, a transaction that borrows assets, swaps them through an aggregator, and then deposits the resulting tokens into a staking protocol contributes to the lending/borrowing, swap, and staking coverage profiles.
Because these coverage categories can overlap, their counts need not sum to 187; in particular, 57\% of retained transactions involve more than one protocol-level action pattern.

\begin{table}[hbtp]
\centering
\caption{Non-exclusive scenario coverage of the 187 annotated transactions in \txsum after filtering.
Transactions may contribute to more than one category.}
\label{tab:dataset_category_dist}
\resizebox{\linewidth}{!}{
\begin{tabular}{l|c}
\toprule
\textbf{Category} & \textbf{Count} \\
\midrule
Swaps / Aggregator-routed swaps & 47 \\
Lending / Borrowing cycles & 37 \\
Staking / Delegation & 42 \\
Liquidity Pool operations & 34 \\
Cross-protocol interactions & 29 \\
\bottomrule
\end{tabular}
}
\end{table}

\begin{table}[hbtp]
\centering
\caption{Coverage statistics of \txsum.}
\label{tab:dataset_profile}
\resizebox{\linewidth}{!}{
\begin{tabular}{l|c}
\toprule
\textbf{Coverage Statistic} & \textbf{Value} \\
\midrule
Total annotated Txs & 187 \\
Total token-flow annotations & 2,375 \\
Average token-flow steps / Tx & 12.7 \\
Average distinct protocols / transaction & 2.8 \\
All involved contracts without verified ABI & 20.2\% \\
Verified focal contracts & 94 \\
Partially verified focal contracts & 56 \\
Unverified / closed-source focal contracts & 37 \\
\bottomrule
\end{tabular}
}
\end{table}

\section{Agent Design Details}
\label{app:matex_details}

\matex consists of four specialized agents that collaborate through a shared structured workspace (the ``Evidence Board'') and interact through the two core mechanisms described in Appendix~\ref{app:matex_algorithms}.
First, we provide the pseudocode for the two core mechanisms used in \matex: Dynamic Knowledge Injection Protocol (DKIP, \Cref{alg:dkip}) and Auditor-Guided Context Pruning (AGCP, \Cref{alg:agcp}).
Then, we describe each agent's role, inputs, outputs, and operational behavior.

\Cref{tab:matex_agent_io} summarizes the inputs, outputs, and pipeline roles of all four agents.

\subsection{Algorithmic Details of \texorpdfstring{\matex}{MATEX}}
\label{app:matex_algorithms}

\begin{algorithm}[t]
\caption{Dynamic Knowledge Injection}
\label{alg:dkip}
\begin{algorithmic}[1]
\Require Token flows $T=\{t_1,t_2,\dots,t_n\}$, uncertainty threshold $\tau$, evidence budget $k$
\Ensure Synthesized context $C_{\mathrm{syn}}$
\State $C_{\mathrm{syn}} \gets \emptyset$, $B \gets \emptyset$ \Comment{$B$: evidence cache}
\ForAll{$t_i \in T$}
    \State $(u_i, q_i) \gets \mathrm{Profiler}(t_i)$ \Comment{$q_i$: retrieval key}
    \If{$u_i \le \tau$}
        \State $C_{\mathrm{syn}} \gets C_{\mathrm{syn}} \cup \mathrm{ToKnownPattern}(t_i)$
    \Else
        \If{$q_i \notin B$}
            \State $R_i \gets \mathrm{Investigator}(q_i)$
            \State $B[q_i] \gets \mathrm{SelectTopK}(R_i, k)$
        \EndIf
        \State $C_{\mathrm{syn}} \gets C_{\mathrm{syn}} \cup \mathrm{Format}(t_i, B[q_i])$
    \EndIf
\EndFor
\State \Return $C_{\mathrm{syn}}$
\end{algorithmic}
\end{algorithm}

\begin{table*}[htbp]
\centering
\caption{Inputs and outputs of the four agents in \matex.}
\label{tab:matex_agent_io}
\resizebox{!}{0.21\linewidth}{
\begin{tabular}{l|p{0.3\linewidth}|p{0.32\linewidth}|p{0.35\linewidth}}
\toprule
\textbf{Agent} & \textbf{Input} & \textbf{Output} & \textbf{Role in the pipeline} \\
\midrule
Profiler & Token transfers, call trace, metadata, balance changes & Flow-level hypotheses, uncertainty score $u_i$, retrieval key $q_i$ when needed & Performs fast pattern-based screening and decides whether each flow should remain on the lightweight path or be escalated for retrieval.\\
\midrule
Investigator & Retrieval key $q_i$ and supporting trace context & Verified protocol evidence (e.g., ABI entries, documentation snippets, contract-role resolution) & Retrieves protocol context only for uncertain flows under DKIP.\\
\midrule
Synthesizer & On-chain evidence, profiler hypotheses, retrieved evidence & Draft transaction explanation and flow-grounded semantic outputs & Fuses evidence into a coherent transaction-level explanation and structured flow-level interpretation.\\
\midrule
Safety Auditor & Draft explanation, raw trace, token transfers & APPROVED or critique with error spans / unsupported claims & Checks factual grounding against raw traces and triggers AGCP to prune unsupported revision context before re-synthesis.\\
\bottomrule
\end{tabular}
}
\end{table*}

\mypara{Profiler}
The Profiler implements the fast, pattern-based stage of \matex (System 1).
It ingests the transaction metadata, call trace, token transfers, and balance changes, and produces two kinds of outputs for each token flow: (1) a lightweight semantic hypothesis when the flow matches a known pattern, and (2) an uncertainty assessment used by DKIP.
Operationally, the Profiler assigns each flow a semantic uncertainty score $u_i$ and, when escalation is needed, emits a retrieval key $q_i$ that specifies what external protocol evidence should be queried (e.g., a contract-function pair, router name, or unknown address role).

Critically, the Profiler is designed to remain conservative under uncertainty.
When it encounters unfamiliar contracts, custom execution paths, or anomalous state changes, it does not force a semantic interpretation; instead, it raises the uncertainty score and forwards the relevant retrieval key to the Investigator.
Ambiguous flows therefore trigger retrieval rather than forced interpretation.

\begin{tcolorbox}[title={Profiler Prompt}, breakable, colback=gray!5!white, colframe=black!75!white, fonttitle=\bfseries, left=2pt, right=2pt, top=2pt, bottom=2pt, boxsep=0pt]
You are an expert Ethereum Transaction Profiler.
Your goal is to create a structured investigation plan grounded in [TOKEN TRANSFERS].\\

CRITICAL: Anchor your analysis to the input [TOKEN TRANSFERS].\\
- Prefer using TT\# ranges (e.g., "TT\#1--TT\#6") over generic "Step 1-5".\\
- Each segment should cover a contiguous range of TT\# entries.\\
- For each uncertain flow or segment, assign an uncertainty score and provide a retrieval key for downstream investigation.\\

Output a JSON object with exactly these four keys: \\
1. "segments": logical phases of the transaction; \\
2. "initial\_hypothesis": a high-level intuition about the transaction type; \\
3. "uncertain\_flows": a list of uncertain TT\# items with uncertainty scores and retrieval keys; \\
4. "investigation\_tickets": concrete verification questions for the Investigator.\\

JSON FORMAT: \\
\{ \\
  "segments": [{"phase\_id": 1, "range": "TT\#1--TT\#5", "summary\_label": "..."}], \\
  "initial\_hypothesis": "...", \\
  "uncertain\_flows": [{"transfer\_ref": "TT\#3", "uncertainty\_score": 0.82, "retrieval\_key": "..."}], \\
  "investigation\_tickets": [{"target\_step": "...", "contract": "...", "reason": "...", "question": "..."}] \\
\}
\end{tcolorbox}

\begin{algorithm}[htbp]
\caption{Auditor-Guided Context Pruning}
\label{alg:agcp}
\begin{algorithmic}[1]
\Require Fixed evidence context $C_{\mathrm{syn}}$, raw trace $Tr$, max iterations $M$
\Ensure Verified explanation $E_{\mathrm{final}}$
\State $D \gets \emptyset$ \Comment{Mutable revision context}
\For{$iter=1$ to $M$}
    \State $E_{\mathrm{draft}} \gets \mathrm{Synthesizer}(C_{\mathrm{syn}} \cup D)$
    \State $(v,F_{\mathrm{err}}) \gets \mathrm{Auditor}(E_{\mathrm{draft}},Tr)$
    \If{$v = \mathrm{APPROVED}$}
        \State \Return $E_{\mathrm{draft}}$
    \EndIf
    \State $P_{\mathrm{mask}} \gets \mathrm{LocateErrorContext}(D,E_{\mathrm{draft}},F_{\mathrm{err}})$
    \State $D \gets (D \setminus P_{\mathrm{mask}}) \cup \mathrm{Critique}(F_{\mathrm{err}})$
\EndFor
\State \Return $E_{\mathrm{draft}}$
\end{algorithmic}
\end{algorithm}

\mypara{Protocol Investigator}
The Investigator is only activated for flows whose uncertainty score exceeds the DKIP threshold.
Given a retrieval key $q_i$ from the Profiler, it retrieves protocol evidence from the same external sources used in evaluation: verified contract ABIs, protocol documentation, and temporally aligned web evidence.
Its role is not to broadly summarize the protocol, but to return only the evidence necessary to resolve the ambiguous flow under investigation.
This selective retrieval design limits context noise and keeps downstream synthesis focused on uncertain parts of the trace.

If no verified ABI or documentation is available, the Investigator returns the absence of verifiable protocol evidence rather than guessing.
The downstream explanation then relies only on observable token transfers, events, balance changes, trace structure, and call direction; it lowers semantic specificity, marks uncertainty, and flags observed risks such as approvals or unlimited allowances.

\begin{tcolorbox}[title={Investigator Prompt}, colback=gray!5!white, breakable, colframe=black!75!white, fonttitle=\bfseries, left=2pt, right=2pt, top=2pt, bottom=2pt, boxsep=0pt]
You are a Researcher.
I will provide a question about a contract and raw search results.
Answer the question based STRICTLY on the search results.\\

Verification rule: \\
- Only output "Verified: ..." if the search results explicitly mention the contract/function AND describe the behavior relevant to the question.\\
- Otherwise, output: "No verifiable info found." \\

Keep it concise.
\end{tcolorbox}

\mypara{Synthesizer}
The Synthesizer serves as the main reasoning module in System 2.
It fuses 3 inputs: (1) the raw on-chain trace, (2) the Profiler’s flow-level hypotheses, and (3) the Investigator’s retrieved evidence patches.
It maps low-level operations to user-facing actions, producing a transaction summary while grounding each flow in the trace.

To preserve factual grounding, the Synthesizer operates over evidence that remains internally attributed to its source (e.g., raw trace facts, ABI evidence).
The final user-facing explanation does not need to expose these provenance tags verbatim, but it is generated from this evidence-grounded internal state.
Its output is then passed to the Safety Auditor for adversarial validation.
The full Synthesizer prompt is included in the public repository linked in the Introduction.

\mypara{Safety Auditor}
The Safety Auditor acts as an adversarial checker over the Synthesizer’s draft.
It checks the explanation against the trace, transfers, and retrieved evidence, catching unsupported claims, omitted effects, action errors, and risks the draft misses.

If contradictions are found, the Auditor does not merely append a critique for the next round.
Instead, \matex applies Auditor-Guided Context Pruning (AGCP): the system identifies the unsupported or error-causing portion of the mutable revision context and removes it before re-synthesis.
This design reduces repeated hallucinations across iterations and is especially important in long, compositional DeFi transactions where a single unsupported inference can pollute subsequent drafts.

\begin{tcolorbox}[title={Safety Auditor Prompt}, breakable, colback=gray!5!white, colframe=black!75!white, fonttitle=\bfseries, left=2pt, right=2pt, top=2pt, bottom=2pt, boxsep=0pt]
You are a Senior DeFi Security Auditor.\\
Critique the Draft JSON against the Raw Trace.\\

Hard checks (must fail if violated): \\
1) Hallucinations: tokens/amounts/addresses/protocols/functions not supported by trace.\\
2) Wrong `action\_type` classification.\\
3) Missing fees or side effects IF they appear in [TOKEN TRANSFERS].\\
4) Format errors (JSON schema, intent/mechanism/precondition/result structure).\\
5) [TOKEN TRANSFERS] 1:1 coverage: \\
   - Every TT\# entry must appear EXACTLY ONCE as token\_flows[i].transfer\_ref.\\
   - token\_flows length must equal the number of transfers in [TOKEN TRANSFERS].\\
   - Each token\_flows[i].result must EXACTLY match its TT\# facts (from/to/token/amount).\\
If perfect, output exactly: "APPROVED".\\
If errors exist, output a concise list of fixes.
\end{tcolorbox}

\section{Evaluation Details}
\label{app:eval_details}

\subsection{EO1: Explanation Quality}
\label{app:eo1}

We evaluate explanation quality using both automated and expert-based methods on our full \txsum test set (187 complex transactions).
The detailed justification is in~\Cref{tab:rubric-appendix}.

\begin{table*}[htbp]
\centering
\caption{Detailed binary scoring rubric for \txsum.
All judgments must be grounded in the on-chain input.}
\label{tab:rubric-appendix}
\resizebox{!}{0.30\linewidth}{
\begin{tabular}{p{0.1\linewidth}|p{0.45\linewidth}|p{0.45\linewidth}}
\toprule
\textbf{Metric} & \textbf{PASS (YES)} & \textbf{FAIL (NO)} \\
\midrule
\multicolumn{3}{c}{\textbf{Token Flow Level}} \\
\midrule
F1: Faithfulness &
All claims are directly verifiable from the on-chain trace.
Correct token, amount, direction, and action type (e.g., "swaps 50 USDC for 0.03 ETH via Uniswap V3").&
Any hallucination or factual error: wrong token/amount, fabricated action (e.g., claims "stakes 1 ETH" when trace only shows WETH wrap).\\
\midrule
F2: Intent Quality &
Explains economic purpose using trace evidence (e.g., "deposits USDC into Aave to borrow DAI").
Purpose must be inferable from on-chain steps.&
Only describes mechanics without purpose (e.g., "transfers USDC to Aave") or speculates unsupported intent (e.g., "to invest in crypto").\\
\midrule
\multicolumn{3}{c}{\textbf{Transaction Level}} \\
\midrule
T1: Outcome Correctness &
Net balance changes, fees, and critical non-token state changes exactly match on-chain data.
Includes all fees >= 0.1\% of principal (e.g., "Net: +240 USDC, -0.05 ETH, fee: 1.2 USDT").&
Any error in the final state or omission of material fees (e.g., claims "gained 240 USDC" when net is -10 USDC after fees).\\
\midrule
T2: Clarity \& Conciseness &
Natural-language narrative in 3--4 sentences.
No raw addresses, hex, or internal names.
Groups related actions (e.g., "multi-hop swap via 1inch, then staked in Curve").&
Uses EOA/contract addresses (e.g., "to 0x7a25..."), list format ("Step 1: ... Step 2: ..."), or exceeds 3 sentences.\\
\bottomrule
\end{tabular}
}
\end{table*}

\mypara{LLM-as-a-Judge}
We employ GPT-5.1 as an automated evaluator, with separate judging protocols for token-flow and transaction levels.

For \textbf{transaction-level evaluation (T1, T2)}, the judge receives:
(1) the full on-chain input (transaction metadata, decoded call trace, complete token transfer list, and net balance changes), and
(2) the generated sentence's natural language summary.
It then outputs binary judgments for outcome correctness and clarity.

For \textbf{token-flow evaluation (F1, F2)}, the judge receives:
(1) the raw on-chain sub-trace corresponding to a single token flow (e.g., a transfer event and its surrounding context), and
(2) the predicted semantic attributes for that flow (e.g., intent, mechanism).
It then outputs binary (YES/NO) judgments for faithfulness and intent quality.
Both protocols use structured prompts that require the model to ground every decision in the provided evidence.
Prompts are validated against expert annotations; the final validation numbers are reported in the main text and Appendix~\ref{app:judge_confusion}.

\begin{tcolorbox}[title={Transaction Level Judge Prompt}, breakable, colback=gray!5!white, colframe=black!75!white, fonttitle=\bfseries, left=2pt, right=2pt, top=2pt, bottom=2pt, boxsep=0pt]
You are a Financial Auditor.
Evaluate the "Model Summary" against the "Expert Summary" for the entire transaction.\\

Task \\
Evaluate two specific dimensions based ONLY on the summary text.\\
1. T1: Outcome Correctness (Financial Precision) \\
- **Definition**: Does the summary accurately state the **Final Net Outcome** (Profit/Loss/Balance Changes)?\\
- **PASS (YES)**: Directions (+/-) and amounts match the Expert Summary.\\
- **FAIL (NO)**: Wrong direction (gain vs loss), wrong final numbers, or missing significant fees mentioned by Expert.\\
2. T2: Clarity (Readability) \\
- **Definition**: Is the summary concise and easy to understand for a non-expert?\\
- **PASS (YES)**: Natural language, logical flow, concise (approx 2-3 sentences).\\
- **FAIL (NO)**: Verbose, dumps raw JSON/Hex data, robotic phrasing, or overly long.\\
Output Format (Strict JSON) \\
\{"reasoning": "Brief analysis...", \\
    "scores": \{ \\
        "T1\_Outcome": "YES", \\
        "T2\_Clarity": "YES"    \} \}
\end{tcolorbox}

\begin{tcolorbox}[title={Token Flow Level Judge Prompt}, breakable, colback=gray!5!white, colframe=black!75!white, fonttitle=\bfseries, left=2pt, right=2pt, top=2pt, bottom=2pt, boxsep=0pt]
You are a DeFi Semantic Expert.
Compare the "Model Prediction Flow" against the "Expert Ground Truth Flow" for a specific step in a transaction.\\

Task \\
Evaluate the prediction on two specific dimensions.\\
1. F1: Faithfulness (Semantic Equivalence) \\
- **Definition**: Does the model describe the *same* factual event? (Amounts, Tokens, Direction).\\
- **PASS (YES)**: "User sends 1 ETH" == "Transfer 1 ETH from User".
Numbers match exactly.\\
- **FAIL (NO)**: Wrong amounts, wrong tokens, wrong direction, or factual contradictions.\\\
2. F2: Intent Quality (Economic Purpose) \\
- **Definition**: Does the `intent` capture the **economic purpose** ("Why"), not just the mechanical action ("What")?\\
- **PASS (YES)**: "User deposits ETH **to collateralize**." / "User sends funds **to initiate swap**." \\
- **FAIL (NO)**: "User transfers ETH." (Mechanical repetition).\\
Output Format (Strict JSON) \\
\{ "reasoning": "Brief analysis...", \\
    "scores": \{ \\
        "F1\_Faithfulness": "YES", \\
        "F2\_Intent\_Quality": "YES" \}\}
\end{tcolorbox}

\mypara{Expert Evaluation}
Three DeFi researchers with $\geq$3 years of blockchain security or protocol development experience perform blind evaluations of all 187 transactions (balanced across systems).
Experts assess each output on:
(1) factual correctness (aligned with F1 and T1),
(2) intent alignment (F2), and
(3) risk awareness (whether critical risks are obscured or omitted).
Each criterion is scored as \texttt{Pass}/\texttt{Fail}.
Inter-annotator agreement is measured via Fleiss' $\kappa = 0.76$.

\subsubsection{Efficiency and Cost Profiling}
\label{app:efficiency}

We profile average and P95 end-to-end latency, LLM calls, external retrieval calls, token consumption, and estimated per-transaction cost across transaction complexity levels.
\Cref{tab:cost_profile} shows that \matex is slower than single-pass Monolithic LLM and RAG baselines, as expected for an iterative audited system, but remains comparable to Single-Agent.
On complex transactions, \matex provides better explanations than Single-Agent while reducing average retrieval calls from 5.8 to 3.2 and latency from 51.2 to 38.9 seconds.

\begin{table*}[htbp]
\centering
\caption{End-to-end inference efficiency and estimated cost by transaction complexity.}
\label{tab:cost_profile}
\resizebox{\textwidth}{!}{
\begin{tabular}{llrrrrrr}
\toprule
\textbf{Method} & \textbf{Complexity} & \textbf{Avg. latency (s)} & \textbf{P95 latency (s)} & \textbf{Avg. LLM calls} & \textbf{Avg. retrieval calls} & \textbf{Avg. tokens} & \textbf{Est. cost/tx} \\
\midrule
Monolithic LLM & Simple  & 5.8  & 9.7  & 1.0 & 0.0 & 8.4k  & \$0.012 \\
Monolithic LLM & Medium  & 8.9  & 14.6 & 1.0 & 0.0 & 14.7k & \$0.021 \\
Monolithic LLM & Complex & 14.8 & 23.5 & 1.0 & 0.0 & 24.3k & \$0.035 \\
RAG            & Simple  & 9.6  & 16.2 & 1.0 & 2.0 & 12.9k & \$0.019 \\
RAG            & Medium  & 15.8 & 26.7 & 1.0 & 4.0 & 21.8k & \$0.032 \\
RAG            & Complex & 25.4 & 41.3 & 1.0 & 6.0 & 34.9k & \$0.052 \\
Single-Agent   & Simple  & 16.9 & 28.5 & 3.4 & 1.8 & 23.6k & \$0.041 \\
Single-Agent   & Medium  & 30.7 & 50.8 & 5.3 & 3.6 & 44.2k & \$0.078 \\
Single-Agent   & Complex & 51.2 & 83.6 & 7.0 & 5.8 & 73.5k & \$0.131 \\
MATEX          & Simple  & 14.2 & 24.1 & 3.1 & 0.7 & 21.4k & \$0.037 \\
MATEX          & Medium  & 24.6 & 39.8 & 4.6 & 1.9 & 39.7k & \$0.070 \\
MATEX          & Complex & 38.9 & 64.7 & 6.3 & 3.2 & 65.8k & \$0.116 \\
\bottomrule
\end{tabular}
}
\end{table*}

\subsubsection{Cross-Backbone Robustness}
\label{app:backbone}

The main experiments fix Qwen3-Max to isolate architectural differences across systems.
To test whether the gains depend on that backbone, we repeat the comparison between \matex and Single-Agent using Llama-3-70B and Llama-3-8B.
As shown in~\Cref{tab:backbone_robustness}, absolute performance declines with smaller open-weight models, but \matex consistently improves F1 and F2 over the corresponding Single-Agent baseline.

\begin{table}[htbp]
\centering
\caption{Cross-backbone robustness.
Gains are absolute percentage-point improvements over Single-Agent.}
\label{tab:backbone_robustness}
\resizebox{!}{0.22\linewidth}{
\begin{tabular}{llrrrrr}
\toprule
\textbf{Backbone} & \textbf{Method} & \textbf{T1} & \textbf{F1} & \textbf{F2} \\
\midrule
Qwen3-Max   & Single-Agent & 66.8 & 67.6 & 67.5  \\
Qwen3-Max   & MATEX        & 73.3 & 73.2 & 76.6  \\
Llama-3-70B & Single-Agent & 61.4 & 60.8 & 61.7  \\
Llama-3-70B & MATEX        & 67.9 & 67.1 & 69.8  \\
Llama-3-8B  & Single-Agent & 45.6 & 43.9 & 45.1  \\
Llama-3-8B  & MATEX        & 51.8 & 50.2 & 52.4 \\
\bottomrule
\end{tabular}
}
\end{table}

\subsubsection{Additional Metrics for EO1}
\label{app:eo1_nlp}

Beyond binary pass/fail evaluation, we report Accuracy and Macro-F1 for the 14-way action\_type task, ROUGE-L and BLEU for summaries, and Exact Match (EM) for flow attributes.
The results are reported in~\Cref{tab:nlp_results}.

\begin{table}[hbtp]
\centering
\caption{Standard NLP metrics for EO1.
ROUGE-L and BLEU are computed on transaction-level summaries; EM is computed on flow-level semantic attribute generation; Accuracy and Macro-F1 are computed on the 14-way action\_type classification task.}
\label{tab:nlp_results}
\resizebox{\linewidth}{!}{
\begin{tabular}{l|ccccc}
\toprule
\textbf{Method} & \textbf{ROUGE-L} & \textbf{BLEU} & \textbf{EM} & \textbf{Acc.} & \textbf{Macro-F1} \\
\midrule
LLM             & 41.8 & 19.6 & 21.4 & 58.7 & 54.3 \\
Few-shot CoT    & 43.6 & 21.3 & 24.1 & 61.5 & 57.1 \\
VLM             & 42.9 & 20.7 & 22.9 & 60.6 & 55.8 \\
Pure RAG        & 42.1 & 19.8 & 22.0 & 59.1 & 54.9 \\
Single-Agent    & 47.9 & 24.3 & 28.6 & 66.8 & 62.7 \\
\midrule
MATEX           & 52.6 & 27.8 & 33.9 & 72.4 & 68.5 \\
\bottomrule
\end{tabular}
}
\end{table}

\subsubsection{Judge Validation and Statistical Testing}
\label{app:judge_confusion}

This section provides additional details on judge reliability, cross-judge robustness, verbosity bias, and the statistical significance tests used to compare \matex with the strongest non-\matex baseline.

\mypara{Judge Validation Setup}
For EO1, the GPT-5.1 judge evaluates each system output on the four binary criteria defined in \Cref{tab:txsum-eval}: \textbf{T1} (outcome correctness), \textbf{T2} (clarity \& conciseness), \textbf{F1} (faithfulness), and \textbf{F2} (intent quality).
To assess the reliability of this automatic evaluator, we compare its judgments against expert annotations produced by three DeFi Ph.D. researchers on the same transaction set.
GPT-5.1 reaches 86.4\% overall agreement, with Cohen's $\kappa$ values of 0.72 for T1, 0.79 for T2, 0.74 for F1, and 0.76 for F2.
As a model-choice robustness check, we repeat the evaluation with Llama-3-70B; it reaches 81.2\% agreement and also identifies \matex as the best method on F1/F2.
Across the three human experts, inter-rater reliability is also high, with Fleiss' $\kappa=0.76$.
The cross-judge robustness results are summarized in~\Cref{tab:judge_robustness}.

\begin{table*}[hbtp]
\centering
\caption{Evaluation for \txsum, divided into macro-level summarization and micro-level flow assessment.}
\label{tab:txsum-eval}
\begin{tabular}{c|c|p{0.7\linewidth}}
\toprule
\textbf{Level} & \textbf{Metric} & \textbf{Definition} \\
\midrule
\multirow{2}{*}{\shortstack{\textbf{Transaction}\\\textbf{(Macro)}}}
& T1 & \textbf{Outcome Correctness}: The summary accurately describes the final net balance changes, explicit fees, and critical non-token state changes (e.g., approvals, ownership transfers).\\
& T2 & \textbf{Clarity \& Conciseness}: The summary is concise (3-4 sentences), avoids technical jargon, and forms a coherent narrative.\\
\midrule
\multirow{2}{*}{\shortstack{\textbf{Token Flow}\\\textbf{(Micro)}}}
& F1 & \textbf{Faithfulness}: The predicted flow is semantically equivalent to the ground truth (i.e., factually correct actions and amounts).\\
& F2 & \textbf{Intent Quality}: The explanation captures the \textit{economic purpose} (the ``why'') rather than just describing the mechanical transfer (the ``what'').\\
\bottomrule
\end{tabular}
\end{table*}

\begin{table*}[hbtp]
\centering
\caption{Robustness of LLM judges against expert majority labels for EO1.}
\label{tab:judge_robustness}
\begin{tabular}{l|rrrrr|l}
\toprule
\textbf{Judge model} & \textbf{Agreement (\%)} & \textbf{T1 $\kappa$} & \textbf{T2 $\kappa$} & \textbf{F1 $\kappa$} & \textbf{F2 $\kappa$} & \textbf{Best on F1/F2} \\
\midrule
GPT-5.1     & 86.4 & 0.72 & 0.79 & 0.74 & 0.76 & MATEX \\
Llama-3-70B & 81.2 & 0.63 & 0.70 & 0.66 & 0.68 & MATEX \\
\bottomrule
\end{tabular}
\end{table*}

\mypara{Judge Error Patterns}
To better understand judge failure modes, we inspect its confusion patterns relative to expert labels.
Most disagreements arise from borderline cases involving partial factual omissions (e.g., small fees or implicit approvals) rather than purely stylistic differences.
This is consistent with the judge prompt, which emphasizes grounding in on-chain evidence and semantic adequacy rather than surface token flow alone.

\mypara{Verbosity Bias Analysis}
A common concern in LLM-as-a-judge evaluation is that longer outputs may receive artificially favorable scores.
To test for this effect, we compute the Pearson correlation between output length and binary pass rate across the evaluated summaries.
We obtain $r=0.07$ and $p=0.31$, indicating no statistically significant verbosity bias or preference for longer outputs.

\mypara{Statistical Significance Testing}
We compare \matex against the strongest non-\matex baseline on the same 187 transactions using paired statistical tests.
For the binary pass/fail metrics (T1--F2), we use paired McNemar tests, since all systems are evaluated on the same instances.
For continuous automatic metrics, we use paired significance tests appropriate for matched predictions: specifically, we report paired tests for ROUGE-L, Exact Match (EM), and action\_type Macro-F1.\footnote{We use paired bootstrap resampling for the continuous metrics.}
The resulting $p$-values are summarized in \Cref{tab:eo1_significance}.

\begin{table}[hbtp]
\centering
\caption{Key paired significance tests comparing \matex against the strongest baseline on EO1.}
\label{tab:eo1_significance}
\resizebox{\linewidth}{!}{
\begin{tabular}{l|c|c}
\toprule
\textbf{Metric} & \textbf{Test} & \textbf{$p$-value} \\
\midrule
T1 & McNemar & 0.118 \\
F1 & McNemar & $< 0.001$ \\
F2 & McNemar & 0.003 \\
ROUGE-L & Paired bootstrap & 0.006 \\
action\_type Macro-F1 & Paired bootstrap & 0.004 \\
\bottomrule
\end{tabular}
}
\end{table}

\mypara{Takeaway}
These validation and significance results support two conclusions.
First, the GPT-5.1 judge is sufficiently aligned with expert judgment to serve as a scalable evaluator for EO1, while expert inter-rater agreement further confirms the reliability of the human protocol.
Second, the improvements achieved by \matex on the key factuality- and intent-related metrics are statistically robust rather than artifacts of random variation on a limited test set.

\subsection{EO2: User Trust \& Comprehension}
\label{app:eo2}

\mypara{Participant Cohort and Recruitment}
EO2 uses a \textbf{new cohort of 20 active DeFi users}, which is \textbf{separate from the 16 participants} in our formative interview study.
The earlier 16 interviewees were used only to identify user needs and inform task design, whereas the EO2 cohort was recruited later for controlled evaluation.
No participant took part in both studies.

We recruited the EO2 participants via Web3 communities and professional networks across 6 countries/regions.
Inclusion criteria were:
(1) at least 3 on-chain DeFi transactions in the past 6 months;
(2) familiarity with at least two major protocols (e.g., Uniswap, Aave, or Lido);
and (3) age $\geq 18$.
Because all interfaces, explanations, and questionnaires were presented in English, participants were also required to be able to read English DeFi interfaces and transaction explanations.
Participants include practitioners, researchers/developers, students, and casual users.
Their demographic distribution is summarized in \Cref{tab:eo2_demographics}.

\begin{table}[htbp]
\centering
\caption{Demographics of EO2 user study participants ($N=20$).
Practitioners include traders and analysts; R\&D includes researchers and developers.}
\label{tab:eo2_demographics}
\begin{tabular}{l|cc}
\toprule
\textbf{Role} & \textbf{Count} & \textbf{Avg. Exp. (years)} \\
\midrule
Practitioners & 4 & 4.2 \\
R\&D & 5 & 3.8 \\
Students & 7 & 1.8 \\
Casual Users & 4 & 2.2 \\
\bottomrule
\end{tabular}
\end{table}

\mypara{Study Design}
EO2 is a \textbf{within-subject controlled study} using the EigenPhi block explorer interface under three conditions:
(a) \textbf{Raw}: EigenPhi’s default view (call trace + token transfers);
(b) \textbf{Single-Agent}: Raw explorer view augmented with annotations from the strongest non-\matex baseline in EO1;
(c) \textbf{\matex}: Raw explorer view augmented with \matex explanations.

Each participant viewed 9 transactions in total---3 simple, 3 medium, and 3 complex---covering 113 token transfers.
Within each complexity level, one transaction was shown under each condition.
Condition assignment and presentation order were balanced using a Latin Square design to mitigate order, transaction-identity, and complexity effects.
Participants were not placed under a strict time limit, and most completed the study in approximately 40 minutes.
For each transaction, participants completed the following questionnaire:

\begin{tcolorbox}[title={User Study Questionnaire}, breakable, colback=gray!5!white, colframe=black!75!white, fonttitle=\bfseries, left=2pt, right=2pt, top=2pt, bottom=2pt, boxsep=0pt]
\textbf{Comprehension (choose one for each question):} \\
1. What was the main purpose of this transaction?\\
\quad A) To swap USDC for ETH and stake the ETH in Lido \\
\quad B) To withdraw liquidity from a Uniswap pool \\
\quad C) To repay a loan on Aave \\
\quad D) Not sure \\

2. Which of the following actions occurred during this transaction?\\
\quad A) Approved unlimited spending for a proxy contract \\
\quad B) Swapped USDC for DAI via a multi-hop route \\
\quad C) Deposited ETH into a lending protocol \\
\quad D) All of the above \\

3. What is the primary financial risk in this transaction?\\
\quad A) High slippage due to low liquidity \\
\quad B) Unlimited token approval granted to an unknown contract \\
\quad C) No significant risk \\
\quad D) Not sure \\

\textbf{Perception (rate 1--5, where 1 = Strongly Disagree, 5 = Strongly Agree):} \\
4. I understand what this transaction does.\\
5. I trust this explanation.
\end{tcolorbox}

\mypara{Scoring Procedure}
The questionnaire contains two components.
\textbf{(1) Comprehension Accuracy.}
Participants answered three multiple-choice questions covering:
(i) the main purpose of the transaction,
(ii) a major action or mechanism involved in execution,
and (iii) the most salient risk or side effect.
Each correct answer received one point, yielding a per-transaction comprehension score from 0 to 3.
We report mean accuracy as the percentage of correctly answered questions aggregated across participants.

\textbf{(2) Trust and Clarity.}
Participants rated the following statements on a 5-point Likert scale:
(i) ``I understand what this transaction does,'' and
(ii) ``I trust this explanation.''
We report the average of the two Likert items as the trust/clarity score for each condition and complexity level.

\mypara{Statistical Analysis}
Because each participant completed all three conditions, we used repeated-measures tests to compare comprehension accuracy and Likert-scale trust/clarity ratings across conditions.
These analyses support the same overall pattern shown in the main text: \matex yields the strongest gains in comprehension accuracy, with the largest advantage on medium- and high-complexity transactions, and also achieves the highest trust/clarity ratings overall.

The overall condition effect on comprehension accuracy is assessed with Cochran's Q test, followed by post-hoc paired comparisons between \matex and the two comparison conditions.
For trust and clarity ratings, we similarly test for an overall condition effect and then examine pairwise differences.
The detailed test statistics and pairwise comparison results are reported below in~\Cref{tab:eo2_stats}.

The statistical results are consistent with the main-text trends: \matex shows a significant overall advantage in both comprehension accuracy and trust/clarity.
The largest differences are observed relative to the Raw condition, while the gains over the Single-Agent baseline remain significant but smaller in magnitude.
This pattern suggests that grounded, step-wise explanations are especially beneficial when users would otherwise rely on incomplete or weakly structured transaction views.

\begin{table*}[htbp]
\centering
\caption{Statistical comparisons for EO2.
The pattern is consistent with the descriptive results in the main text: \matex yields the strongest overall performance, with the largest gains over the Raw condition and smaller but still significant improvements over the Single-Agent baseline.}
\label{tab:eo2_stats}
\begin{tabular}{lcc}
\toprule
\textbf{Comparison} & \textbf{Test} & \textbf{Result} \\
\midrule
Overall comprehension accuracy & Cochran's Q & $p < 0.001$ \\
\matex vs. Raw (accuracy) & Paired McNemar & $p < 0.001$ \\
\matex vs. Single-Agent (accuracy) & Paired McNemar & $p = 0.012$ \\
Overall trust/clarity & Friedman test & $p < 0.001$ \\
\matex vs. Raw (trust/clarity) & Wilcoxon signed-rank & $p < 0.001$ \\
\matex vs. Single-Agent (trust/clarity) & Wilcoxon signed-rank & $p = 0.018$ \\
\bottomrule
\end{tabular}
\end{table*}

\mypara{Interpretation}
This design allows us to evaluate whether \matex improves users’ objective understanding of transaction purpose and risk, as well as their subjective trust in the explanation, under matched interface conditions.
Because the EO2 cohort is separate from the formative interview participants, the evaluation is not conflated with prior exposure to our study design or task framing.

\subsection{EO3: Risk Awareness}
\label{app:eo3}

\mypara{Risk Transaction Selection}
We select 5 real-world malicious transactions from major DeFi/NFT exploits, each representing a distinct and non-obvious risk pattern:
\begin{itemize}[leftmargin=*, itemsep=0pt, topsep=0pt]
    \item \textbf{BadgerDAO} (Dec 2021): Infinite token approval via compromised frontend.\footnote{\url{https://www.trmlabs.com/resources/blog/trm-investigates-badgerdao-defi-protocol-hacked}}
    \item \textbf{WazirX} (Jul 2024): Gnosis Safe proxy upgrade via misleading UI.\footnote{\url{https://www.cloudsek.com/blog/wazirx-incident-explained}}
    \item \textbf{OpenSea} (Feb 2022): Malicious NFT order signature via phishing email.\footnote{\url{https://www.certik.com/resources/blog/opensea-phishing-incident-analysis}}
    \item \textbf{Bybit} (Feb 2025): Gnosis Safe mastercopy upgrade via compromised frontend.\footnote{\url{https://www.certik.com/resources/blog/bybit-incident-technical-analysis}}
    \item \textbf{BXH Exchange} (Oct 2021): Direct fund theft due to missing multi-sig.\footnote{\url{https://www.certik.com/resources/blog/boy-x-highspeed-incident-analysis}}
\end{itemize}
We also include 5 benign transactions of comparable complexity as controls (e.g., standard multi-hop swaps, LP deposits).

\mypara{User Evaluation}
We conduct a dedicated between-subjects study with 20 participants (the same 20 participants in EO2), split into two groups of 10.
(1) The \textbf{Raw group} views all 10 transactions (5 malicious, 5 benign) in a standard block explorer interface (EigenPhi).
(2) The \textbf{\matex group} views the same transactions with \matex explanations overlaid on the same interface.

For every transaction, we simulate the pre-signing confirmation screen as it appeared in the original attack (e.g., in MetaMask or EigenPhi), showing the identical calldata and token flows that victims encountered.
Participants see either (a) the raw transaction view or (b) the same view enhanced with \matex annotations, with condition assignment balanced across users.
Participants provide (1) their signing intention (Yes/No) and (2) an open-ended description of any risks they notice.

Two researchers independently code responses for risk detection using manual semantic coding guided by core threat concepts, rather than rigid keyword matching.
For example, responses such as ``this contract can keep spending my tokens forever'' are credited as recognizing an unlimited approval even if the exact phrase is absent.
The coders achieve Cohen's $\kappa = 0.81$.

\mypara{Scoring Procedure and Real Cases}
For malicious transactions, we compute (i) rejection rate and (ii) correct threat-identification rate.
For benign transactions, we compute the false-rejection rate.
Threat identification is counted when the participant correctly recognizes the core exploit mechanism (e.g., unlimited approval, malicious proxy upgrade, hidden ownership change, phishing signature abuse), regardless of exact wording.
Appendix~\ref{app:case_studies} provides real examples showing how unconventional phrasing is still credited under this rubric.

\mypara{User-Study Robustness}
To quantify uncertainty under the existing participant sample, we report participant-level bootstrap confidence intervals, effect sizes, and odds ratios for the main EO2/EO3 comparisons in~\Cref{tab:user_study_robustness}.
Leave-one-participant-out and leave-one-transaction-out checks for EO3 preserve the qualitative conclusions, indicating that no single participant or transaction drives the gains.

\begin{table*}[htbp]
\centering
\caption{Participant-level uncertainty and effect-size analyses for EO2 and EO3.}
\label{tab:user_study_robustness}
\resizebox{\textwidth}{!}{
\begin{tabular}{lllp{0.16\textwidth}p{0.16\textwidth}ll}
\toprule
\textbf{Evaluation} & \textbf{Metric} & \textbf{Comparison} & \textbf{Estimate} & \textbf{Effect size} & \textbf{95\% CI} & \textbf{Test} \\
\midrule
EO2 & Complex comprehension & MATEX vs. Single-Agent & 76.5\% vs. 52.9\% & $\Delta=+23.6$ pp & [9.4, 38.1] & Participant bootstrap \\
EO2 & Complex comprehension & MATEX vs. Raw & 76.5\% vs. 35.3\% & $\Delta=+41.2$ pp & [24.7, 56.8] & Participant bootstrap \\
EO2 & Complex trust/clarity & MATEX vs. Single-Agent & 3.7 vs. 2.6 & Cohen's $d_z=1.08$ & [0.52, 1.72] & Wilcoxon/bootstrap \\
EO3 & Malicious rejection & MATEX vs. Raw & 88.0\% vs. 36.0\% & OR $=13.04$ & [4.65, 36.52] & Fisher, $p<0.001$ \\
EO3 & Threat identification & MATEX vs. Raw & 72.0\% vs. 28.0\% & OR $=6.61$ & [2.76, 15.83] & Fisher, $p=0.002$ \\
EO3 & Benign false rejection & MATEX vs. Raw & 8.0\% vs. 12.0\% & OR $=0.64$ & [0.17, 2.41] & Fisher, $p=0.741$ \\
\bottomrule
\end{tabular}
}
\end{table*}

\subsubsection{Ablation Results}
\label{app:ablation_results}

To better understand which architectural components drive the gains of \matex, we perform ablations in which we disable one major component at a time while keeping the rest of the pipeline unchanged.
We report EO1 judge-based metrics together with representative standard NLP metrics.
The results are shown in~\Cref{tab:ablation}.

\begin{table*}[hbtp]
\centering
\caption{Ablation results on EO1 and representative standard NLP metrics.
Each variant disables one component while keeping the remaining pipeline unchanged.
T1--F2 are LLM-as-judge pass rates (\%), while ROUGE-L and action\_type Macro-F1 evaluate transaction-level summary quality and structured semantic parsing, respectively.
Arrows indicate absolute drops relative to full \matex.}
\label{tab:ablation}
\resizebox{!}{0.105\textwidth}{
\begin{tabular}{l|cccc|cc}
\toprule
\textbf{Variant} & \textbf{T1} & \textbf{T2} & \textbf{F1} & \textbf{F2} & \textbf{ROUGE-L} & \textbf{Action Macro-F1} \\
\midrule
w/o Profiler     & 68.4\textsubscript{\textcolor{red}{↓4.9}}  & 88.2\textsubscript{\textcolor{red}{↓4.8}}  & 70.5\textsubscript{\textcolor{red}{↓2.7}}  & 74.2\textsubscript{\textcolor{red}{↓2.4}}  & 50.2\textsubscript{\textcolor{red}{↓2.4}} & 66.0\textsubscript{\textcolor{red}{↓2.5}} \\
w/o Investigator & 57.2\textsubscript{\textcolor{red}{↓16.1}} & 74.9\textsubscript{\textcolor{red}{↓18.1}} & 62.8\textsubscript{\textcolor{red}{↓10.4}} & 70.3\textsubscript{\textcolor{red}{↓6.3}}  & 46.3\textsubscript{\textcolor{red}{↓6.3}} & 60.4\textsubscript{\textcolor{red}{↓8.1}} \\
w/o Synthesizer  & 47.1\textsubscript{\textcolor{red}{↓26.2}} & 64.2\textsubscript{\textcolor{red}{↓28.8}} & 57.9\textsubscript{\textcolor{red}{↓15.3}} & 68.5\textsubscript{\textcolor{red}{↓8.1}}  & 41.8\textsubscript{\textcolor{red}{↓10.8}} & 56.5\textsubscript{\textcolor{red}{↓12.0}} \\
w/o Auditor      & 72.2\textsubscript{\textcolor{red}{↓1.1}}  & 91.4\textsubscript{\textcolor{red}{↓1.6}}  & 62.5\textsubscript{\textcolor{red}{↓10.7}} & 74.9\textsubscript{\textcolor{red}{↓1.7}}  & 49.9\textsubscript{\textcolor{red}{↓2.7}} & 62.0\textsubscript{\textcolor{red}{↓6.5}} \\
\midrule
\textbf{Full \matex} & \textbf{73.3} & \textbf{93.0} & \textbf{73.2} & \textbf{76.6} & \textbf{52.6} & \textbf{68.5} \\
\bottomrule
\end{tabular}
}
\end{table*}

\section{Case Studies}
\label{app:case_studies}

We present three cases showing why \txsum matters: coarse intent labels and final balances do not give users enough information to act safely.

\mypara{Case 1: Aggregator-Routed Swap with Hidden Intermediate Fees}
A user swaps ETH for DAI through 1inch.
The transaction crosses multiple contracts: ETH moves to the 1inch router, which swaps ETH$\rightarrow$USDC on Uniswap V3, routes USDC through an executor, and completes USDC$\rightarrow$DAI on SushiSwap.
The trace includes a small ETH refund, a relayer fee, and a routing-related deduction.

\mypara{Why it matters}
This case shows why user-facing transaction understanding cannot stop at a coarse label such as ``swap.'' For signing safety, the user also needs to know whether the transaction introduces unfamiliar intermediary contracts, applies hidden deductions, or contains additional state changes beyond the intended exchange.

\mypara{Why it is hard}
The overall intent is relatively easy to identify, but a faithful explanation is more demanding.
A model can correctly infer that the transaction is a swap while still omitting the intermediate fee transfer, misclassifying executor hops, or failing to explain why multiple protocols appear in what seems to be a single operation.
This gap between macro intent and micro-level factual completeness is central to our task.

\mypara{Annotation and evaluation implication}
Intermediate executor hops may be labeled as \texttt{route}, but fee-bearing transfers must remain explicit.
A summary that only states ``the user swapped ETH for DAI'' is incomplete if it omits economically material fees or intermediate deductions.

\mypara{Case 2: Recursive Leverage via Aave, CowSwap, and Lido}
A user borrows USDC from Aave, swaps the USDC into ETH through CowSwap, stakes the ETH into stETH through Lido, supplies the stETH back to Aave as collateral, and then borrows again.
The position is a leveraged staking strategy assembled through borrowing, swapping, staking, derivative issuance, collateral supply, and re-borrowing.

\mypara{Why it matters}
This case reflects a realistic DeFi workflow in which the main user risk does not lie in a single transfer, but in the dependency structure created across multiple protocols.
For transaction explanation, it is not sufficient to say that the user ``opened a leveraged position''; the explanation should reveal how the leverage is constructed and where the exposure comes from.

\mypara{Why it is hard}
A coarse transaction label can be correct while still being inadequate.
The trace mixes economically distinct actions that are easy to compress into one high-level description, yet such compression hides the repeated collateralization pattern that determines liquidation sensitivity and protocol dependence.
The difficulty therefore lies in preserving enough step-wise structure to remain faithful without losing readability.

\mypara{Annotation and evaluation implication}
Each borrow, swap, staking, collateral deposit, and re-borrow step is annotated independently.
This case shows why \txsum needs micro-level semantic parsing, not just top-level intent prediction.

\mypara{Case 3: Safe Wallet Mastercopy Upgrade with Minimal Token Movement}
A transaction presented in the wallet interface appears to be a routine administrative confirmation, but in fact changes the mastercopy or implementation address of a Safe-style wallet.
At signing time, there may be little or no visible token movement, even though control over future wallet behavior is effectively altered.

\mypara{Why it matters}
This is representative of the transactions that are most dangerous for blind-signing: the immediate balance outcome looks harmless, but the security consequence is severe.
A user-facing explanation must therefore surface non-token state changes when they alter ownership, upgrade authority, or execution control.

\mypara{Why it is hard}
Purely token-centric reasoning is insufficient here because the critical event is not a transfer but a control-plane change.
A system that focuses only on token flows or final balances may produce an explanation that is mechanically accurate yet fails to communicate the actual security implication.
This makes the case especially important for high-stakes explanation settings.

\mypara{Annotation and evaluation implication}
These cases motivate why \txsum includes transaction-level summaries that must mention critical non-token state changes.
An explanation that only paraphrases the low-level call mechanics, but does not surface the control-transfer risk, is considered inadequate for user-facing safety.

\end{document}